\pdfoutput=1 
\documentclass[a4paper,11pt]{article}
\PassOptionsToPackage{table}{xcolor}

\usepackage{jheppub} 

\usepackage[T1]{fontenc} 
\usepackage{tangocolors}

\usepackage{tikz}
\usetikzlibrary{arrows,snakes,backgrounds}
\usetikzlibrary{decorations.pathmorphing,backgrounds,shapes,arrows,shadows}
\tikzset{zigzag/.style={decorate,decoration=zigzag}}
\tikzset{snake it/.style={decorate, decoration=snake}}
\makeatletter
\def\@hex@@Hex#1%
 {\if a#1A\else \if b#1B\else \if c#1C\else \if d#1D\else
  \if e#1E\else \if f#1F\else #1\fi\fi\fi\fi\fi\fi \@hex@Hex}
\makeatother

\usepackage[all]{xy}
\usepackage[percent]{overpic}
\usepackage{slashed}
\usepackage{wrapfig}
\usepackage{tabu}
\usepackage{diagbox}
\usepackage{mathrsfs,amsmath,amssymb,amsthm,amsfonts,tikz,graphicx,accents,hyperref, color}
\usepackage{dsfont,epiolmec, latexsym, stmaryrd, comment}
\usepackage{slashed,ccaption}
\usepackage{mathrsfs, calligra}
\usepackage{leftidx}
\usepackage{import}
\usepackage{multirow}
\usepackage{amsfonts}
\usepackage{pifont}
\usepackage{tabularx}
\usepackage{cancel}
\usepackage[utf8]{inputenc}
\usetikzlibrary{intersections,calc}
\usepackage{ifthen}
\usepackage{amsmath}
\usepackage{cancel}
\usepackage{caption}

\usepackage{array}
\hypersetup{ linktoc=all,
    colorlinks, linkcolor={palatinateblue},
    citecolor={brightpink}, urlcolor={amaranth}
}

\graphicspath{{Images/}}

\renewcommand{\d}[1]{\ensuremath{\operatorname{d}\!{#1}}}

\def\sideremark#1{\ifvmode\leavevmode\fi\vadjust{\vbox to0pt{\vss
 \hbox to 0pt{\hskip\hsize\hskip1em
 \vbox{\hsize2cm\tiny\raggedright\pretolerance10000
 \noindent #1\hfill}\hss}\vbox to8pt{\vfil}\vss}}}%
\DeclareSymbolFont{extraup}{U}{zavm}{m}{n}
\DeclareMathSymbol{\varheart}{\mathalpha}{extraup}{86}
\DeclareMathSymbol{\vardiamond}{\mathalpha}{extraup}{87}
\makeatletter
\renewcommand*{\@fnsymbol}[1]{\ensuremath{\ifcase#1\or \clubsuit \or \vardiamond \or \varheart\or
    \spadesuit\or \mathparagraph\or \|\or **\or \dagger\dagger
    \or \ddagger\ddagger \else\@ctrerr\fi}}
\makeatother

\definecolor{rosy}{RGB}{230,235,252}
\definecolor{myframetitle}{RGB}{90,89,170}
\definecolor{myblocktitle}{RGB}{140,185,249}
\definecolor{mytitle}{RGB}{10,80,26}

\definecolor{darkgreen}{RGB}{27,130,45}
\definecolor{darkblue}{rgb}{0,0,0.3}
\definecolor{darkred}{rgb}{0.7,0,0}

\definecolor{light gray}{RGB}{220,220,220}
\definecolor{dark purple}{RGB}{108,0,217}
\definecolor{pink}{RGB}{190,20,100}
\definecolor{orang}{RGB}{193,63,0}
\definecolor{green}{RGB}{11,98,17}
\definecolor{darkpink}{RGB}{153,0,76}
\definecolor{bluegreen}{RGB}{0,102,102}
\definecolor{greenlagan}{RGB}{0,102,0}
\definecolor{redgreen}{RGB}{102,102,0}
\definecolor{Redgreen}{RGB}{153,76,0}
\definecolor{vividviolet}{rgb}{0.62, 0.0, 1.0}
\definecolor{amaranth}{rgb}{0.9, 0.17, 0.31}
\definecolor{palatinateblue}{rgb}{0.15, 0.23, 0.89}
\definecolor{brightpink}{rgb}{1.0, 0.0, 0.5}
\definecolor{cornflowerblue}{rgb}{0.39, 0.58, 0.93}
\definecolor{deepcarminepink}{rgb}{0.94, 0.19, 0.22}
\definecolor{radicalred}{rgb}{1.0, 0.21, 0.37}

\newcommand\bc[1]{\boldsymbol{\mathcal{#1}}}

\DeclareFontFamily{OT1}{rsfs}{}

\DeclareFontShape{OT1}{rsfs}{m}{n}{ <-7> rsfs5 <7-10> rsfs7 <10->rsfs10}{} 

\DeclareMathAlphabet{\mycal}{OT1}{rsfs}{m}{n}

\newcommand{\be}{\begin{equation}}
\newcommand{\ee}{\end{equation}}
\newcommand{\bea}{\begin{eqnarray}}
\newcommand{\eea}{\end{eqnarray}}

\makeatletter \@addtoreset{equation}{section}

\begin{document}

\newcommand{\mytitle}{\centerline{\LARGE{Null Surface Thermodynamics in Topologically Massive Gravity}}}\vskip 3mm 

\title{{\mytitle}}
\author[]{ Vahid Taghiloo}

\affiliation{Department of Physics, Institute for Advanced Studies in Basic Sciences (IASBS),
P.O. Box 45137-66731, Zanjan, Iran, and}
\affiliation{ School of Physics, Institute for Research in Fundamental
Sciences (IPM), P.O.Box 19395-5531, Tehran, Iran}

\emailAdd{
v.taghiloo@iasbs.ac.ir
}

\abstract{We study three dimensional topologically massive gravity (TMG) in presence of a generic codimension one null boundary. The existence of the boundary is accounted for by enlarging the Hilbert space of the theory by degrees of freedom which only reside at the boundary, the \textit{boundary degrees of freedom}. The solution phase space of this theory in addition to  
bulk massive chiral gravitons of the TMG, involves boundary modes which are labeled by surface charges associated with large diffeomorphisms. We show boundary degrees of freedom obey a local thermodynamic description over the solution phase space, \textit{null surface thermodynamics}, described by a local version of the first law, a local Gibbs-Duhem equation, and local zeroth law.  Due to the expansion of the boundary and also the passage of the bulk mode through the boundary, our null surface thermodynamics describes an open boundary system that is generically out of thermal equilibrium.
}
\maketitle

\section{Introduction}
Formulating gravity theories in presence of boundaries brings in new degrees of freedom (d.o.f) which only reside at the boundary: boundary d.o.f. 
Therefore the existence of boundaries requires enlarging the solution space of the theory in such a way it captures these new boundary d.o.f.
The first natural question in this regard is how can we describe these boundary modes? 
To answer this question we revisit more carefully gauge theories or diffeomorphism invariant theories of gravity in presence of boundaries. 

Gauge theories  enjoy local symmetries and we usually treat them as redundancies in description of theory. But in presence of boundaries a part of these transformations, large gauge transformations/diffeomorphisms, can become physical. They are large in the sense that they act non-trivially on the boundary (Cauchy) data. 
Different boundary data correspond to different solutions, so these transformations by definition act as nontrivial maps on the solution space of the theory. In this sense large diffeomorphisms are symmetries and 
boundary data can be labelled by their associated charges. We now have all the ingredients to answer the above mentioned question: We use the surface charges associated with large diffeomorphisms to label our desired boundary d.o.f. Dynamics of these boundary d.o.f is constrained by the “refined equivalence principle” \cite{Sheikh-Jabbari:2016lzm} which also takes into account the features and properties of the boundary.

Motivated by these, we study gravitational theories on spacetimes with a null boundary. This boundary can be an arbitrary null surface  in spacetime {and is not necessarily horizon of a black hole} or asymptotic infinity of an asymptotic flat space time. This has been the research program pursued in some recent {works} \cite{Adami:2020amw, Grumiller:2020vvv, Adami:2020ugu, Adami:2021sko} and in particular in \cite{Adami:2021nnf, Grumiller:2019fmp, Donnay:2015abr, Donnay:2016ejv, Afshar:2016wfy, Afshar:2016uax,Afshar:2016kjj, Hopfmuller:2016scf, Hopfmuller:2018fni, Donnay:2018ckb, Chandrasekaran:2018aop, Chandrasekaran:2019ewn, Chandrasekaran:2020wwn, Ciambelli:2021vnn,Freidel:2021fxf, Freidel:2021cbc}. 
In reference \cite{Adami:2021nnf}, $D$-dimensional Einstein gravity in presence of a null boundary was studied. In this case, the solution phase space of the theory was constructed and its symmetries and corresponding charges were analyzed. The solution phase space of this theory is parameterized by $D(D-3)$ infalling and outgoing bulk propagating gravitons and also $D$ boundary d.o.f. It was shown in  \cite{Adami:2021kvx} that these boundary d.o.f along with the bulk modes describe an open thermodynamic system with local laws. These local laws of thermodynamics account for the dynamics of the part of spacetime behind the boundary.

There is another way to view the null surface thermodynamics: One can interpret the standard first law of black hole thermodynamics as a relation between hard charges (i.e. mass, angular momentum, and ...). The content of the standard first law is actually the conservation of energy. It states how the black hole's hard charges should be changed through a perturbation. From the soft hair proposal \cite{Haco:2018ske,Hawking:2016msc} we know the black holes carry infinite number of soft charges. So we expect through a perturbation which carries the soft hair, the black hole soft hair should be changed in such a way that the total amount of the soft hair remain intact. Now the question is whether we have a similar first law for the soft charges. Because of the conservation of soft charges we expect to exist such a soft version of the first law. 

As pointed out in \cite{Adami:2021kvx,Sheikh-Jabbari:2022xix}, the local laws of thermodynamics which describe the dynamics of boundary d.o.f are a direct consequence of diffeomorphism invariance of the action. One can ask how  crucial is the diffeomorphism invariance to get the thermodynamic description for the boundary d.o.f. In order to answer this question we go beyond the gravitational theories with covariant action and consider a \textit{semi-covariant} action. Here semi-covariant means equations of motion are generally covariant while the action  under general diffeomorphisms  transforms up to a total derivative term.  An important example of semi-covariant theories is three dimensional Topologically Massive Gravity (TMG) \cite{deser1982three,deser2000topologically}. The TMG action  involves the gravitational Chern-Simon term which is not diffeomorphism invariant but transforms up to a total derivative term under a general coordinate transformation. A distinctive feature of this theory is that it has a massive propagating graviton. Therefore, another motivation to consider TMG  is to explore the role of  massive gravitons on the boundary thermodynamics.

Null boundary analyses for topologically massive gravity has been carried out in  \cite{Adami:2021sko}.\footnote{Various aspects of topologically massive gravity has been extensively studied in the recent literature,  see e.g. \cite{Macias:2005pm, Chow:2009km, Chow:2009vt, chow:2019ucq, Gurses:2010sm, Gurses:2011fv, Ertl:2010dh, Deser:2009er, Garbarz:2008qn, Carlip:2008eq, Aliev:1995cf, Nutku:1993eb,Anninos:2008fx,sachs:2011xa, Bonora:2011gz, Detournay:2012ug, Compere:2009zj, Bouchareb:2007yx, Moussa:2003fc,Detournay:2015ysa}.} In this case, the solution space of the theory is described by four functions. Three of them describe the boundary d.o.f and the fourth represents the massive chiral propagating mode of TMG. In this paper, we show these boundary d.o.f in presence of this chiral mode describe an open thermodynamic system. We present a local version of the first law and Gibbs-Duhem equation. The form of these equations is the same as \cite{Adami:2021kvx}, but the local thermodynamic quantities also receive contributions from the Chern-Simon term in the action of this theory. We also discuss the local zeroth law for boundary thermodynamics. As we will see the local zeroth law yields the  Heisenberg $\oplus$ Vir  algebra among the thermodynamic quantities, where the central charge of the Viraroso algebra is proportional to the gravitational Chern-Simons coupling, as the one obtained in \cite{Adami:2021sko}. To perform our analyses we first construct  solution phase space of the theory perturbatively around the null boundary and analyze the symmetries over the solution space. We recognize two class of solutions, the vanishing Cotton tensor (VCT)  and the non-vanishing Cotton tensor (NVCT) cases. The former coincides with what we have in the absence of the Chern-Simons term, analyzed in \cite{Adami:2021sko}. 

This paper is organized as follows. In section \ref{sec:review}, we review the solution phase space and boundary symmetries for the TMG \cite{Adami:2021sko}. Sections \ref{sec: Null surface thermodynamics (VCT)} and \ref{sec: Null surface thermodynamics (NVCT)} contain our main results for the local thermodynamics of TMG for expanding null hypersurfaces in the VCT and NVCT cases. They involve a local version of the first law, a local Gibbs-Duhem equation, and a statement for the local zeroth law. In section \ref{Non-expanding case}, we consider the thermodynamics of non-expanding null boundaries. In section \ref{sec:discussion}, we discuss further our results and conclude with an outlook.
\section{Null Surface Solution Phase Space, A Review}\label{sec:review}
Topologically massive gravity (TMG), with negative cosmological constant $\Lambda= -1/\ell^2$, is described by the action \cite{deser1982three,deser2000topologically},
\begin{equation}\label{TMG-action}
I[g]=\frac{1}{16\pi G}\int \mathrm{d}^{3}x\ L[g],\qquad L[g]:=\sqrt{-g} \left(R+\frac{2}{\ell^2}+\frac{1}{\mu}L_{\text{\tiny CS}} \right)
\end{equation}
where $R$ is Ricci scalar and $L_{\text{\tiny CS}}$ is the gravitational Chern-Simons term,
\begin{equation}
L_{\text{\tiny CS}}= \frac{1}{2}\epsilon^{\mu\nu\rho}\left(\Gamma^{\alpha}_{\mu\beta}\partial_{\nu}\Gamma^{\beta}_{\rho\alpha}+\frac{2}{3}\Gamma^{\alpha}_{\mu\beta}\Gamma^{\beta}_{\nu\gamma}\Gamma^{\gamma}_{\rho\alpha}\right)
\end{equation}
with $\epsilon_{\mu\nu\lambda}$ being the Levi-Civita tensor which in our conventions  $\sqrt{-g}\epsilon^{vr\phi}=1$, and $\Gamma^\alpha_{\mu\nu}$ is the Christoffel symbol. This action has three parameters of dimension length, $G, \ell$ and the Chern-Simons coupling $1/\mu$. Equations of motion are a system of third order partial differential equations which may also be  written as \cite{Adami:2021sko}
\begin{equation}\label{extra-condition}
    \mathcal{E}^{\mu}_{\nu} := 
    \mathcal{D}^{\mu}{}_{\beta} \, \mathcal{T}^{\beta}{}_{\nu}=0\,, 
\end{equation}
where
\begin{equation}\label{GRE}
    \mathcal{T}_{\mu \nu}:=R_{\mu \nu}+ \frac{2}{\ell^2} g_{\mu \nu}\, ,\qquad 
    \mathcal{D}^{\mu}{}_{\nu} = \delta^{\mu}{}_{\nu}+ \frac{1}{\mu} \epsilon^{\mu \alpha}{}_{\nu} \nabla_{\alpha}\, .
\end{equation}
We split the solution space of this theory in two different categories: $\mathcal{T}_{\mu \nu}=0$ and $\mathcal{T}_{\mu \nu}\neq 0$. They are respectively called vanishing Cotton tensor (VCT) and non-vanishing Cotton tensor (NVCT) sectors \cite{Adami:2021sko}. The first class only involves the Einstein solutions but the second class contains solutions that do not appear in the solution space of 3-dimensional Einstein gravity. The latter one contains a massive propagating mode which due to the appearance of the Levi-Civita tensor we call, \textit{chiral massive news}.
We adopt Gaussian null-type coordinate system 
\begin{equation}
    g_{rr}=0\, , \hspace{1 cm}  g_{r\phi}=0\, , \hspace{1 cm} \partial_{r}g_{rv}=0\, 
\end{equation}
in which $v,r,\phi$ are respectively advanced time, radial and angular coordinates.
By this kind of gauge fixing, the three dimensional line-element is given by \cite{Adami:2020ugu,Adami:2021nnf, Adami:2021sko}
\begin{equation}\label{G-N-T-metric}
    \d s^2=  -V \d v^2 + 2 \eta \d v \d r + {\cal R}^2 \left( \d \phi + U \d v \right)^2\, 
\end{equation}
where $V,\mathcal{R},$ and $U$ are generic functions on spacetime and $\eta$ depends only on $v,\phi$. In this coordinate system, we take $r=0$ to be a null surface, $V(r=0)=0$, and we denote it by $\mathcal{N}$. By assuming the Taylor expandability of the line element around our desired null boundary ($r=0$), we do the following expansions
\begin{subequations}\label{r-expansion-3d'}
    \begin{align}
       V(v,r,\phi) =&  -\eta\left(\Gamma+\frac{\mathcal{D}_{v}\eta}{\eta}\right)r +r^2 V_2+\mathcal{O}(r^3)\, , \\
       U(v,r,\phi) =& \,\mathcal{U}- r \frac{\eta\mathcal{J}}{\Omega^3} 
       +\mathcal{O}(r^2)\, , \\
       {\cal R}^2(v,r,\phi) =&\, \Omega^2(1 -2r\eta\Theta_{n}) 
       +\mathcal{O}(r^2)\, .
    \end{align}
\end{subequations}
All functions which appear in these expansions are generic functions of null boundary coordinates, $v$ and $\phi$. These functions have nice geometrical meanings. For example, we could think about $\Gamma$, $\mathcal{U}$, $\mathcal{J}$, $\Theta_{n},$ and $\Omega$ as the surface gravity, velocity aspect of the boundary, angular momentum aspect, expansion, and the area density of the null boundary respectively \cite{Adami:2021sko}.
For latter convenience, we introduce the differential operators $\mathcal{D}_v$ and $\mathcal{L}_\mathcal{U}$ which their action on a codimension one function $O_w (v,\phi)$ of weight $w$ is defined through \cite{Adami:2022ktn, Adami:2021nnf}
\begin{subequations}
    \begin{align}
    \mathcal{D}_v O_w := &\partial_v O_w - \mathcal{L}_\mathcal{U} O_w \, ,\\
    \mathcal{L}_{\mathcal{U}} O_w := & \mathcal{U} \partial_\phi O_w + w O_w \partial_\phi \mathcal{U} \, , 
    \end{align}
\end{subequations}
where $\mathcal{U}$ is a function of weight $-1$. Weights of  different functions can be found in Table \ref{Table-1}.
\begin{table}[t]
\centering
\begin{tabular}{ |l|l| }
  \hline
  $w= -1$  & $\mathcal{U}$ ,  $Y$ , $\hat{Y}$ \\
  $w= 0$  & $\eta$  , $T$ , $W$ , $\mathcal{T}_{ll}$ , $\Theta_{l}$ , $\Theta_{n}$ , $\kappa$ , $\Gamma$ , $N$ , $\mathcal{E}$ , $V_{2}$ , $\mathcal{E}_{ll}$ , $\hat{T}$ , $\hat{W}$ , $\partial_{v}$ , $\mathcal{D}_{v}$\\
  $w= 1$  & $\Omega$ , $\omega$ , $\mathcal{T}_{l\phi}$ , $\bar{\Omega}$ , $\mathcal{E}_{l\phi}$ , $\bar{\Omega}_{\text{\tiny{NE}}}$ , $\partial_{\phi}$ \\
  $w= 2$  & $\mathcal{T}_{\phi\phi}$ , $\mathcal{J}$ , $\bar{\mathcal{J}}$ , $\bar{\mathcal{J}}_{\text{\tiny{NVCT}}}$ , $\bar{\mathcal{J}}_{\text{\tiny{NE}}}$\\
  \hline
\end{tabular}
\caption{Weight $w$ for various quantities defined and used in this section.}\label{Table-1}
\label{table:weight}
\end{table}
To describe the geometry of the null boundary, we define the following two null vector fields
\begin{equation}\label{gennullbndryl}
    l :=  l_\mu \d x^\mu= -\frac{1}{2} V \d v  + \eta \d r , \qquad
    n :=  n_\mu \d x^\mu= -\d v \, ,
\end{equation}
they satisfy $l^2=n^2=0$ and $l.n=-1$. The vector field $l^\mu \partial_\mu = \partial_v - \mathcal{U} \partial_\phi + \mathcal{O}(r)$ is the generator of the null surface
\begin{equation}\label{kappa-l}
  l \cdot \nabla l^{\mu}:= \kappa\, l^{\mu} \hspace{.2 cm} \text{on $\mathcal{N}$}, \hspace{1 cm} \kappa = -\frac{\Gamma}{2}+ \frac{\mathcal{D}_v \eta}{2\eta} \, ,
\end{equation}
here $\kappa$ is the non-affinity of null boundary generators. In the rest of the work, the on-shell divergence-free and traceless tensor $\mathcal{T}_{\mu \nu}$ \eqref{GRE} will be of great relevance. 
The components $\mathcal{T}_{ll}=l^\mu l^\nu \mathcal{T}_{\mu\nu}$, $\mathcal{T}_{l\phi}=l^\mu  \mathcal{T}_{\mu\phi},$ and $\mathcal{T}_{\phi\phi}= \mathcal{T}_{\phi\phi}$ computed at $r=0$ are given by 
\begin{subequations}\label{Tmunu-components}
\begin{align}\label{Tll}
 & \mathcal{T}_{ll}=-\mathcal{D}_{v}\Theta_{l}+\kappa\Theta_{l}-\Theta_{l}^2\, ,\\
   & \mathcal{T}_{l\phi}= \mathcal{D}_v \omega + \Theta_{l}\omega- \partial_\phi \kappa \, ,\\
    &\mathcal{T}_{\phi\phi}=2\Omega^2\left[
   \mathcal{D}_v \Theta_{n} + \Theta_{n}( \kappa+\Theta_{l}) -\frac{1}{\Omega}\partial_\phi \left( \frac{\omega}{\Omega}\right) -\frac{\omega^2}{\Omega^2}+\frac{1}{\ell^2}\right] \, .
\end{align}
\end{subequations}
Equations of motion \eqref{extra-condition} lead to
\begin{subequations}
    \begin{align}
        \mathcal{E}:= & {\cal E}^{\mu}_\mu=- \frac{2 V_2}{\eta^2}+\frac{3\mathcal{J}^2}{2 \Omega^4}+\frac{2}{\ell^2} -\frac{\mathcal{J} \partial_\phi \eta}{\Omega^3 \eta}-\frac{(\partial_\phi \eta)^2}{2\eta^2 \Omega^2} +\frac{2\mathcal{T}_{\phi \phi}}{\Omega^2}=0\, ,\\
        \mathcal{E}_{ll}:=& l^\mu l^\nu \mathcal{E}_{\mu \nu}=\mathcal{T}_{ll}-\frac{1}{\mu \Omega}  \biggl[\mathcal{D}_v \mathcal{T}_{l\phi}+\omega \mathcal{T}_{ll} +(\Theta_{l}-\kappa) \mathcal{T}_{l\phi} -\partial_\phi \mathcal{T}_{ll}-\frac{\mathcal{U}\Omega^2}{2}\mathcal{D}_{v}\mathcal{E}\biggr]=0 \, ,\\
        \mathcal{E}_{l\phi} :=&  l^{\mu}\mathcal{E}_{\mu\phi} =\mathcal{T}_{l\phi}
    -\frac{1}{\mu \Omega} \biggl[  \mathcal{D}_v \mathcal{T}_{\phi \phi} - \omega \mathcal{T}_{l \phi} - \frac{1}{2}\Theta_{l}\mathcal{T}_{\phi \phi} + \Omega^{2} \Theta_{n}\mathcal{T}_{ll}-\Omega \partial_\phi \left( \frac{\mathcal{T}_{l\phi}}{\Omega}\right)-\frac{1}{2}\Omega^2\Theta_{l}  \mathcal{E}\biggr]=0\, ,
    \end{align}
\end{subequations}
where
\begin{equation}
   \Theta_{{l}} :=  \frac{\mathcal{D}_{v}\Omega}{\Omega}\, , \hspace{1 cm}   \omega :=\frac{1}{2} \left( \frac{\mathcal{J}}{\Omega}+ \frac{\partial_\phi \eta}{\eta}\right)\, 
\end{equation}
are respectively expansion and twist fields.
For later convenience, we also introduce\footnote{It is important to emphasize that $\mathcal{P}$ does not have a well-defined weight. We define its derivative as
\begin{equation*}
    \mathcal{D}_{v}\mathcal{P}:=\partial_{v}\mathcal{P}-\mathcal{L}_{\mathcal{U}}\mathcal{P}, \hspace{1 cm}  \mathcal{L}_{\mathcal{U}}\mathcal{P}:=\mathcal{U}\partial_{\phi}\mathcal{P}-2\partial_{\phi}\mathcal{U}\, .
\end{equation*}}
\begin{equation}
    \mathcal{P}:= \ln \left(\frac{\eta}{\Omega^2\Theta_{l}^{2}}\right)\, 
\end{equation}
here we have assumed $\Theta_{l}\neq 0$. For the non-expanding case, we will define another related quantity. We have summarized the definition of different symbols in Table \ref{Table-2}.
\begin{table}[ht]
\centering
\begin{tabular}{||c c||} 
 \hline
Symbol & Description \\ [0.5ex] 
 \hline\hline
  VCT & Vanishing Cotton Tensor \\
  NVCT & Non-Vanishing Cotton Tensor\\
  NE & Non-Expanding \\
  $\mu$ & Coupling constant of Chern-Simon term\\
  $\mathcal{N}$ & Null boundary \\
  $\mathcal{U}$ & Angular velocity of null boundary\\
  $\kappa$ &  Non-affinity of null boundary generators \\
  $\Theta_{l}$ & Expansion of null vector $l$ \\ 
  $\Theta_{n}$ & Expansion of null vector $n$ \\ 
  $\omega$ &  Twist field \\ 
  $T$ & Generator of supertranslations \\
  $W$ & Generator of superscaling in the radial direction \\
  $Y$ & Generator of superrotations \\
  $T_{_\mathcal{N}}$ & Local temperature \\
  $\Omega$ & Area density of null boundary\\ 
  $\bar{\Omega}$ & Entropy aspect in VCT case\\ 
  $\bar{\Omega}_{\text{\tiny{NE}}}$ & Entropy aspect for non-expanding null boundaries \\
  $\bar{\mathcal{J}}$ & Angular momentum aspect in VCT \\
  $\bar{\mathcal{J}}_{\text{\tiny{NVCT}}}$ & Angular momentum aspect in NVCT \\
  $\bar{\mathcal{J}}_{\text{\tiny{NE}}}$ & Angular momentum aspect for non-expanding null boundaries \\
  $N$ & Chiral massive news \\
  $\mathcal{D}_{v}$ & Comoving derivative along the null boundary\\
  $\approx$ & On-shell equality \\[1ex] 
 \hline
\end{tabular}
\caption{List of frequently occurring symbols.}\label{Table-2}
\label{table:2}
\end{table}
\paragraph{Solution phase space.} 
The solution phase space is parameterized by four functions, $\{\Omega,\eta,\mathcal{J}; \mathcal{T}_{ll}\}$. The first three functions label the boundary d.o.f while the last one captures the information about the chiral massive news. In the VCT case, we lose the bulk propagating mode, and hence, in this case, the dynamic of the theory only arises from the boundary \cite{Adami:2021sko}.
\paragraph{Null boundary symmetry (NBS).} 
The vector field \cite{Adami:2020ugu,Adami:2021nnf, Adami:2021sko}
\begin{equation}\label{null-boundary-sym-gen}
    \xi =T\, \partial_v + r \left(\mathcal{D}_v T -  W  \right) \partial_r + \left( Y -r \frac{\eta}{\Omega^2}  \partial_{\phi} T\right)\partial_{\phi} + \mathcal{O}(r^2)\, ,
\end{equation}
preserves the form of metric \eqref{G-N-T-metric} and hence rotates us in the solution space. Here symmetry generators, $T$, $Y,$ and $W$ are generic functions of null boundary coordinates. They respectively generate the supertranslations in $v$-direction, superrotations in $\phi$-direction, and superscaling in $r$-direction. It is worth to emphasis these three generators are in one to one correspondence with three labels of the boundary d.o.f $\{\Omega,\eta,\mathcal{J}\}$. 
\paragraph{Null boundary symmetry algebra.} 
The vector fields \eqref{null-boundary-sym-gen} make an algebra. Due to the field dependency of \eqref{null-boundary-sym-gen}, we should use the adjusted Lie bracket to read the algebra \cite{Barnich:2011mi, Compere:2015knw}
\begin{equation}\label{3d-NBS-KV-algebra}
    [\xi(  T_1, W_1, Y_1), \xi( T_2,  W_2, Y_2)]_{{\text{adj. bracket}}}=\xi(  T_{12}, W_{12}, Y_{12})
\end{equation}
where 
\begin{subequations}\label{W12-T12-Y12}
\begin{align}
    &T_{12}=\left(T_{1}\partial_{v}+Y_{1}\partial_{\phi}\right)T_{2}-(1\leftrightarrow 2),\\
    &W_{12}=\left(T_{1}\partial_{v}+Y_{1}\partial_{\phi}\right)W_{2}-(1\leftrightarrow 2),\\
    &Y_{12}=\left(T_{1}\partial_{v}+Y_{1}\partial_{\phi}\right)Y_{2}-(1\leftrightarrow 2).
\end{align}
\end{subequations}
This is a Diff$(\mathcal{N})\ \inplus $ Weyl$(\mathcal{N})$ algebra \cite{Adami:2020ugu,Adami:2021sko}. The Diff part of this algebra is parameterized by supertranslations and superrotations and the Weyl part is also described by the generator of superscaling in the $r$ direction.
\paragraph{Field variations of chemical potentials.} 
Under the action of $\xi$, we get
\begin{subequations}\label{delta-fields}
    \begin{align}
        \delta_{\xi}\Gamma &=\mathcal{D}_{v}(W+\Gamma T)+\hat{Y}\partial_{\phi}\Gamma\,,\label{delta-Gamma} \\ 
        \delta_\xi \mathcal{U} &=\mathcal{D}_{v}\hat{Y}\,,\label{delta-UA}\\
        \delta_\xi {\cal P} &=(T\mathcal{D}_{v}+\mathcal{L}_{\hat{Y}})\mathcal{P}-W\, ,\label{delta-P}\\
        \delta_{\xi} \Omega &=(T\mathcal{D}_{v}+\mathcal{L}_{\hat{Y}})\Omega\, ,
   \end{align}
\end{subequations}     
where $\hat{Y}=Y+\mathcal{U}T$. We will interpret $\{\Gamma, \mathcal{U}; \mathcal{P},\Omega\}$ as chemical potentials in thermodynamic equations. The distinctive common feature of these quantities is that they do not depend on the underlying theory and are only geometrically determined.
As the last point, it is worth to emphasis that these field variations are totally off-shell.
\section{Null Surface Thermodynamic, VCT Case} \label{sec: Null surface thermodynamics (VCT)}
In this section, we consider the vanishing Cotton tensor (VCT) sector of the solution phase space of the theory. In all of this section, the on-shell equality $\approx$ means we apply the VCT equations of motion
\begin{equation}\label{EOM-VCT-1}
\mathcal{T}_{ll}=\mathcal{T}_{l\phi}=\mathcal{T}_{\phi\phi}=0\,.
\end{equation}
The first two equations $\mathcal{T}_{ll}=0$ and $\mathcal{T}_{l\phi}=0$ are the standard Raychaudhuri and Damour \cite{PhysRevD.18.3598}
equations respectively. We introduce two further quantities which will play important roles in our null surface thermodynamic 
 \begin{equation}\label{Charges-VCT-1}
    \begin{split}
        \bar{\Omega}=&\Omega+\frac{1}{\mu}\left[\frac{\mathcal{J}}{2\Omega}+\frac{\partial_{\phi}\mathcal{P}}{2}+\frac{\partial_{\phi}\Omega}{\Omega}\right]\, ,\\
        \bar{\mathcal{J}}=&\mathcal{J}+\frac{1}{\mu}\left[\left(\frac{\mathcal{J}}{2\Omega}\right)^2+\left(\frac{\Omega}{\ell}\right)^2+2\Omega^2\Theta_{l}\Theta_{n}-\left(\frac{\partial_{\phi}\mathcal{P}}{2}\right)^2-\frac{\partial_{\phi}\Omega\partial_{\phi}\mathcal{P}}{\Omega} +2\left(\frac{\partial_{\phi}\Omega}{\Omega}\right)^2-\frac{2\partial_{\phi}^2\Omega}{\Omega}\right] \, . \end{split}
\end{equation}
From now on, the barred notion will be used for the quantities which take corrections from the Chern-Simon term in TMG action \eqref{TMG-action}. In other words, these barred quantities are defined in such a way that they reduce to the corresponding unbarred quantities (Einstein counterparts) for $\mu \rightarrow \infty$.
\paragraph{Symplectic potential.} 
We start by calculating the Lee-Wald symplectic potential \cite{Lee:1990nz} for the VCT case. Up to total derivative terms w.r.t $\phi$ we get
\begin{equation}
  16\pi G\,  \Theta_{\text{\tiny{LW}}}^{r}= -\bar{\mathcal{J}}\delta\mathcal{U}+\bar{\Omega}\delta \Gamma+\mathcal{D}_{v}\bar{\Omega}\,\delta\mathcal{P}-\frac{1}{\mu}\partial_{v}\mathcal{A} +\frac{1}{\mu}\delta \mathcal{B}
\end{equation}
where the last two terms are a total derivative term w.r.t $v$ coordinate and a total variation on solution phase space which their explicit form is given by
\begin{equation}
    \begin{split}
        \mathcal{A}=& \delta \omega+\Omega^2 \Theta_{n} \, \delta \mathcal{U} +\frac{\delta \Omega \partial_\phi \Omega}{2 \Omega^2}+\frac{\delta \eta \partial_\phi \eta}{4 \eta^2}  +\frac{\delta \Omega}{\Omega} \partial_\phi \mathcal{P}+\frac{1}{4}\delta\mathcal{P}\partial_{\phi}\mathcal{P}\, ,\\
        \mathcal{B}=&\partial_{v}\omega+\frac{1}{2}\Omega^{2}\Theta_{n}\mathcal{D}_{v}\mathcal{U}-\frac{\mathcal{J}}{4\Omega}(\Gamma+2\Theta_{l})-\frac{\partial_{\phi}\Omega}{\Omega}(\Gamma+\Theta_{l})+\frac{\partial_{\phi}\mathcal{P}}{4}(-\Gamma+2\Theta_{l}+\partial_{\phi}\mathcal{U})\\
        & -\frac{\partial_{\phi}\mathcal{U}}{2}\left(\mathcal{U}\Omega^2\Theta_{n}+\frac{\mathcal{J}}{2\Omega}+\frac{\partial_{\phi}\Theta_{l}}{\Theta_{l}}-\frac{3\partial_{\phi}\Omega}{\Omega}\right)\, .
    \end{split}
\end{equation}
To remove these terms, we use the freedoms/ambiguities in the covariant phase space method \cite{Wald:1999wa}. To do so, we introduce the following $W$ and $Y$ terms
\begin{equation}\label{W-Y-term-VCT}
    W^{\mu}=\frac{\sqrt{-g}\mathcal{B}}{\mu\, \Omega}n^{\mu}\, , \hspace{1 cm} Y^{\mu \nu}[\delta g ; g]= -\frac{\sqrt{-g}}{16 \pi G \mu} \, \epsilon^{\mu \nu \lambda} B_\lambda [\delta g ; g]\, 
\end{equation}
with $B_\lambda$ depending on the (variations) of  metric and Christoffel symbols and the quantity $\mathcal{P}$
\begin{equation}\label{B}
    B_\lambda [\delta g ; g] =\frac{1}{4} \,\Gamma^\alpha_{\lambda \beta} \delta g^\beta_\alpha -\, n_\alpha l^\beta \delta \Gamma^{\alpha}_{\lambda \beta} {+ \frac{\delta \Omega}{ \Omega} \partial_\lambda \mathcal{P}}+\frac{1}{4}\delta\mathcal{P}\partial_{\lambda}\mathcal{P} \,,
\end{equation}
It should be noted these freedoms are proportional to $1/\mu$ and vanish in Einstein limit $\mu\rightarrow\infty$.
By adding these $Y$ and $W$ terms to the Lee-Wald symplectic potential, we find
\begin{equation}
  16\pi G\,  \Theta^{r}= -\bar{\mathcal{J}}\delta\mathcal{U}+\bar{\Omega}\delta \Gamma+\mathcal{D}_{v}\bar{\Omega}\,\delta\mathcal{P}\, .
\end{equation}
From now on, we add these freedoms \eqref{W-Y-term-VCT} to any Lee-Wald quantities in the VCT case and drop their LW index. It is worth emphasizing that these kinds of $Y$-terms \eqref{W-Y-term-VCT} were used to obtain integrable surface charges for TMG in VCT case \cite{Adami:2021sko}.
\paragraph{Symplectic form.} 
One can compute the Lee-Wald pre-symplectic form \cite{Lee:1990nz} over the  set of geometries \eqref{G-N-T-metric}. After the addition of $Y$-term \eqref{W-Y-term-VCT}, it yields
\begin{equation}\label{presymplectic-VCT}
    \begin{split}
     \Omega
        =  \frac{1}{16 \pi G} \int_{\mathcal{N}} \d v \d{}\phi \,\left[\delta\mathcal{U}\wedge \delta\bar{\mathcal{J}}-\delta\Gamma\wedge \delta\bar{\Omega}-\delta\mathcal{P}\wedge\delta(\mathcal{D}_{v}\bar{\Omega})\right].
    \end{split}
\end{equation}
This pre-symplectic form has a nice property: similar to the Einstein gravity in 3 dimensions, it involves three conjugate pairs. In each pair, one accepts corrections from the Chern-Simon term and the other one does not. In this regard, we interpret unchanged quantities, $\{\Gamma, \mathcal{U}; \mathcal{P}\}$, as chemical potentials, which do not depend on the underlying theory, they are only geometrical quantities\footnote{For an exception see \cite{Hajian:2020dcq}.}. We consider $\{\bar{\Omega}, \bar{\mathcal{J}};\mathcal{D}_{v}\bar{\Omega}\}$ as their corresponding thermodynamic conjugate charges which carry information about the underlying theory \eqref{Charges-VCT-1}. In this sense, they are actually dynamic variables. It is important to emphasize the pre-symplectic form \eqref{presymplectic-VCT} involves off-shell quantities. These thermodynamic quantities are subject to the VCT equations of motion \eqref{charge-eom-VCT-1}.

\paragraph{Equations of motion in terms of charges.} 
VCT equations of motion \eqref{EOM-VCT-1} in terms of thermodynamics variables yield
\begin{equation}\label{charge-eom-VCT-1}
    \mathcal{D}_{v}\bar{\Omega}\approx\Theta_{l}(2\Omega-\bar{\Omega})\, , \hspace{1 cm} \mathcal{D}_{v}\mathcal{P}\approx \Gamma\, , \hspace{1 cm} \mathcal{D}_{v}\bar{\mathcal{J}}\approx -\bar{\Omega}\partial_{\phi}\Gamma-\mathcal{D}_{v}\bar{\Omega}\partial_{\phi}\mathcal{P}-2\partial_{\phi}\left(\mathcal{D}_{v}\bar{\Omega}+\frac{\partial_{\phi}^{2}\mathcal{U}}{\mu}\right)\, .
\end{equation}
The second and third equations are the Raychaudhuri and Damour equations which now have been written in terms of thermodynamics variables. They capture the time evolution of expansion and angular momentum.
\paragraph{Charge variations.} 
Under the action of large diffeomorphisms \eqref{null-boundary-sym-gen}, the thermodynamic charges transform as
\begin{subequations}\label{delta-charges}
    \begin{align}      
         \delta_{\xi}\bar{\Omega}\approx & (T\mathcal{D}_{v}+\mathcal{L}_{\hat{Y}})\bar{\Omega}\, ,\\
       \delta_\xi (\mathcal{D}_{v}\bar{\Omega}) \approx & \mathcal{D}_{v}(T\mathcal{D}_{v}\bar{\Omega})+\mathcal{L}_{\hat{Y}}(\mathcal{D}_{v}\bar{\Omega})\, ,\\
      \delta_{\xi}\bar{\mathcal{J}}\approx & (T\mathcal{D}_{v}+\mathcal{L}_{\hat{Y}})\bar{\mathcal{J}}+\bar{\Omega}(\partial_{\phi}W+\Gamma\partial_{\phi}T)-2\mathcal{D}_{v}\bar{\Omega}\partial_{\phi}T+\frac{2}{\mu}\left(T\partial_{\phi}^{3}\mathcal{U}-\partial_{\phi}^{3}\hat{Y}\right)\, .
\end{align}
\end{subequations}
\subsection{Surface Charge Variation} 
One can compute the charge variation associated with large diffeomorphisms for topologically massive gravity by using an extension of the covariant phase space method \cite{Iyer:1994ys,Tachikawa:2006sz, Kim:2013cor,Lee:1990nz,Wald:1999wa}.\footnote{As we mentioned the Chern-Simon term in the TMG Lagrangian \eqref{TMG-action} makes this action to be semi-covariant. So, we need to revisit the standard Noether-Wald method \cite{Iyer:1994ys} for computing surface charges, see e.g. \cite{Tachikawa:2006sz, Kim:2013cor,Adami:2021sko}.} Explicit calculations lead to the following expression for the charge variation
\begin{equation}\label{appchargedeltaoffshell}
\begin{split}
    \slashed{\delta} Q(\xi) \approx  \frac{1}{16 \pi G}   \int_0^{2 \pi} \d \phi \biggl\{  W\delta\bar{\Omega}+Y\delta\bar{\mathcal{J}}+T \left(\mathcal{U}\delta\bar{\mathcal{J}}-\Gamma\delta\bar{\Omega}+\mathcal{D}_{v}\bar{\Omega}\,\delta\mathcal{P} \right)\biggr\}\, .
    \end{split}
\end{equation}
To get this result, we have used the VCT equations of motion \eqref{EOM-VCT-1} and we have also added $Y$-term \eqref{W-Y-term-VCT} into the Lee-Wald surface charge formula. Obviously, in this slicing of the solution phase space ($\delta{T}=\delta{Y}=\delta{W}=0$), the charge variation is not integrable. So, to obtain well-defined labels for the boundary d.o.f, we need to split it into the integrable and non-integrable (flux) parts
\begin{equation}\label{integrable-charge-Y-VCT}
    Q^{\text{I}}(\xi)=\frac{1}{16 \pi G} \int_0^{2 \pi} \d \phi \left[ W \bar{\Omega}+Y \bar{\mathcal{J}}+T(\mathcal{U}\bar{\mathcal{J}}-\Gamma \bar{\Omega})\right]\, ,
\end{equation}
and
\begin{equation}\label{Non-Integrable-Charge-Y-VCT}
    \mathcal{F}(\xi)=\frac{1}{16 \pi G} \int_0^{2 \pi} \d \phi \, T \left(-\bar{\mathcal{J}}\delta\mathcal{U}+\bar{\Omega}\delta \Gamma+\mathcal{D}_{v}\bar{\Omega}\,\delta\mathcal{P}\right)\, .
\end{equation}
By using the modified bracket \cite{Barnich:2011mi} for this splitting of the charge variation,
\begin{subequations}\label{BT-Bracket}
    \begin{align}
   &\delta_{\xi_{2}}Q^{\text{I}}_{\xi_{1}} := \left\{Q^{\text{I}}_{{\xi_{1}}},Q^{\text{I}}_{{\xi_{2}}}\right\}_{\text{\tiny BT}} {-} F_{\xi_{2}}(\delta_{\xi_{1}}g) \label{2d-BT-Bracket-01}\\
     &\left\{Q^{\text{I}}_{\xi_{1}},Q^{\text{I}}_{\xi_{2}}\right\}_{\text{\tiny BT}} =\, Q^{\text{I}}_{[\xi_{1},\xi_{2}]_{{\text{adj. bracket}}}}+K_{\xi_1,\xi_2}
     \label{2d-BT-Bracket-02}
\end{align}
\end{subequations}
we obtain an algebra the same as the NBS algebra \eqref{W12-T12-Y12} with the following central extension term
\begin{equation}\label{central-charge-1}
     K_{\xi_1,\xi_2}=\frac{1}{16 \pi G\,\mu} \int_0^{2 \pi} \d \phi(\hat{Y}_{2}\partial_{\phi}^3\hat{Y}_{1}-\hat{Y}_{1}\partial_{\phi}^3\hat{Y}_{2})\, .
\end{equation}
This result is consistent with the representation theorem in the covariant phase space method \cite{Lee:1990nz, Iyer:1994ys}.
Due to the existence of $\mathcal{U}$ in the definition of $\hat{Y}$, this central extension term is actually field dependent. This central charge is related to the gravitational anomaly of the presumed dual 2d CFT \cite{Kraus:2005vz, Kraus:2005zm, Solodukhin:2005ah}. This matches with the usual statement that central charges are “anomalies” for conserved charges.

One can read the zero mode charges from the full non-integrable form of the charge variation \eqref{charge-variation-Y-1}
\begin{equation}\label{Charges-zero-mode-full-VCT}
\begin{split}
        Q({-r\partial_r}) &:= \frac{\mathbf{\bar{S}}}{4\pi}=\frac{1}{16\pi G}  \int_0^{2 \pi} \d \phi\, \bar{\Omega},     \\ 
        Q({\partial_{\phi}}) &:= \mathbf{\bar{J}}=\frac{1}{16\pi G}  \int_0^{2 \pi} \d \phi\,\bar{\mathcal{J}},\\
       \slashed{\delta} Q({\partial_v}) &:= \slashed{\delta}\mathbf{\bar{H}}=\frac{1}{16\pi G}  \int_0^{2 \pi} \d \phi\, \left(\mathcal{U}\delta\bar{\mathcal{J}}-\Gamma\delta\bar{\Omega}+\mathcal{D}_{v}\bar{\Omega}\,\delta\mathcal{P} \right) \,.
\end{split}
\end{equation}
They have obtained by putting $\xi=-r\partial_r$ , $\xi=\partial_{\phi}$ and $\xi=\partial_v$ in \eqref{charge-variation-Y-1} respectively. One can do the same job with the integrable part of the charge variation \eqref{integrable-charge-Y-VCT},
\begin{equation}\label{Charges-zero-mode-integrable-VCT}
\begin{split}
        Q^{\text{I}}({-r\partial_r}) &:= \frac{\mathbf{\bar{S}}}{4\pi}=\frac{1}{16\pi G}  \int_0^{2 \pi} \d \phi\, \bar{\Omega},     \\ 
        Q^{\text{I}}({\partial_{\phi}}) &:= \mathbf{\bar{J}}=\frac{1}{16\pi G}  \int_0^{2 \pi} \d \phi\, \bar{\mathcal{J}},\\
       Q^{\text{I}}({\partial_v}) &:= \mathbf{\bar{E}}=\frac{1}{16\pi G}  \int_0^{2 \pi} \d \phi\, \left(\mathcal{U}\bar{\mathcal{J}}-\Gamma \bar{\Omega}\right) \,.
\end{split}
\end{equation}
The first zero mode, $\mathbf{\bar{S}}$, is equal to the Wald entropy \cite{Wald:1993nt, Iyer:1994ys} and two other ones correspond to the angular momentum and energy respectively.
\paragraph{Balance equation.}
Now, we consider the conservation of our surface charges. Due to the non-integrability and explicit $v$-dependence of the charge variation, we do not expect our charges to be conserved. In this case, we should consider the balance or generalized charge conservation equation \cite{Barnich:2011mi,Adami:2020amw}
\begin{equation}\label{balance-equation-VCT}
\boxed{    \frac{\d {}}{\d v} Q_{\xi}^{{\text{\tiny{I}}}}\approx-\mathcal{F}_{\partial_{v}}(\delta_{\xi}g;g)+{K}_{\xi,\partial_{v}}\, .}
\end{equation}
This equation relates the time evolution of the integrable part of the surface charges to the non-integrable (flux) and anomalies of the charges. In other words, it states there exist two sources for the non-conservation of the surface charges: flux and anomalies of charges.
\subsection{Null Boundary Thermodynamical Phase Space, VCT Case}\label{sec:Thermodynamical phase space-VCT}
As we discussed one can use the boundary charges to label our boundary d.o.f. In this regard, the VCT part of the solution phase space is parameterized by three boundary d.o.f, $\{\bar{\Omega}, \bar{\mathcal{J}};\mathcal{P}\}$. These boundary d.o.f are led to the following thermodynamic picture \cite{Adami:2021kvx}.
\begin{enumerate}
\item[I.] The null boundary solution space for the VCT case consists of the following two parts:
\begin{enumerate}
    \item[I.1)] thermodynamic sector: parametrized by $(\Gamma, {\cal U})$ and conjugate charges $(\bar{\Omega}, \bar{\mathcal{J}})$. They are subject to the equations of motion \eqref{charge-eom-VCT-1}.
    \item[I.2)] ${\cal P}$, which only appears in  the flux \eqref{Non-Integrable-Charge-Y-VCT} and not in the integrable charge \eqref{integrable-charge-Y-VCT}. As we mentioned, this quantity is not affected by the Chern-Simon term. The thermodynamic conjugate associated with ${\cal P}$ is equal to the time derivative of the entropy aspect.
\end{enumerate}
    \item[II.] The VCT part of the solution phase space does not involve any bulk modes. So, our boundary thermodynamics is only affected by boundary effects. In this regard, $\mathcal{P}$ is a boundary effect which takes our boundary system out of thermal equilibrium (OTE).
    \item[III.] The time derivatives of entropy aspect $\mathcal{D}_{v}\bar{\Omega}$ and area density $\mathcal{D}_{v}{\Omega}$  measure the OTE from the bulk and boundary viewpoints respectively. In the Einstein gravity because we have $\bar{\Omega}=\Omega$, so $\mathcal{D}_{v}{\Omega}$ (or expansion $\Theta_{l}$) is a measure of OTE from both bulk and boundary viewpoints.
\end{enumerate}

In the rest, we clarify this picture by going through the equations.
\subsection{Local First Law at Null Boundary}\label{subsec:first-law}
One can read the standard first law of black hole thermodynamics for stationary black holes (e.g. BTZ black holes in TMG) from \eqref{charge-variation-Y-1} by putting $W=Y=0$ and $T=1$,
\begin{equation}\label{global first law}
{\delta} \bc{\bar{H}}_{0}=T_{0} \delta\bc{\bar{S}}_{0} +{\cal U}_{0}\delta\bc{\bar{J}}_{0}
\end{equation}
where $\bc{\bar{H}}_{0}$ on the left hand side is the energy of the black hole and on the right hand, the first pair (heat term) involves the temperature, $T_{0}$, and the entropy, $\bc{\bar{S}}_{0}$, and the second term (work term) involves ${\cal U}_{0}$ and $\bc{\bar{J}}_{0}$ which are angular velocity and angular momentum respectively. We refer to \eqref{global first law} as the \textit{global} first law. 

In this work we generalize this global equation in three different ways: 1) We present a local version of the first law which holds in each point of the null boundary. 2) We generalize this equation such that it captures the OTE effects. 
As we will see, in the VCT case the expansion of the null boundary makes our boundary system to be OTE. In the NVCT case in addition to this effect, the passage of bulk news through the boundary is another source of OTE in the boundary system.
3) This local equation holds on any null surface and in this description, black holes do not play any key roles. It should be emphasized that all of these analyses are direct consequences of diffeomorphism invariance (at the level of equations of motion).

Now, we present a local version of the first law which characterizes the dynamics of boundary d.o.f. To obtain the \textit{local} first law we start from the full form of the charge variation \eqref{charge-variation-Y-1} and calculate it for $W=Y=0$ and $T=\delta(\phi-\phi')$. Then, it yields the following local equation 
\begin{equation}\label{local-first-law-VCT}
\boxed{\slashed{\delta} \bc{\bar{H}}=T_{_{\mathcal{N}}} \delta\bc{\bar{S}} +{\cal U}\delta\bc{\bar{J}} +\mathcal{D}_{v}\bc{\bar{S}}\,\delta \bc{P}}
\end{equation}
where
\begin{equation}
    \bc{\bar{\mathcal{S}}}=\frac{\bar{\Omega}}{4 G}\, ,\hspace{1 cm} \bc{P}=\frac{\mathcal{P}}{4\pi}\, ,  \hspace{1 cm} T_{_{\mathcal{N}}}:= -\frac{\Gamma}{4\pi}\approx- \mathcal{D}_{v}\bc{P}\, , \hspace{1 cm} \bc{\bar{\mathcal{J}}}=\frac{\bar{\mathcal{J}}}{16\pi G}  \,.
\end{equation}
This equation is a local equation in boundary coordinates, $v$ and $\phi$. Let us look at the right hand side of this equation: the first term is the heat term, $T_{_{\mathcal{N}}}$ is the local temperature which is proportional to the surface gravity and its thermodynamic conjugate is the entropy aspect, $\bc{\bar{\mathcal{S}}}$. The second term is a work term, $\mathcal{U}$ is the angular velocity aspect of the boundary and its conjugate denotes the angular momentum aspect, $\bc{\bar{\mathcal{J}}}$. We interpret the last term in \eqref{local-first-law-VCT} as an entropy production term. Because of the expansion of the null surface, the left hand side unlike the global first law of thermodynamics \eqref{global first law} is not a total variation on the thermodynamic phase space. The integrated version of the local first law \eqref{local-first-law-VCT} for stationary spacetimes reduces to the usual global first law \eqref{global first law}.
\subsection{Local Gibbs-Duhem Equation at Null Boundary}\label{subsec:GD-relation}
In this subsection, we study the global and local Gibbs-Duhem equations. From the integrable part of the charge variation \eqref{integrable-charge-Y-VCT}, we find the standard Gibbs-Duhem equation by substituting $W=Y=0$ and $T=1$,
\begin{equation}\label{global GD}
\bc{\bar{E}}_{0}=T_{0} \bc{\bar{S}}_{0} +{\mathcal{U}}_{0} \bc{\bar{J}}_{0}
\end{equation}
Similar to the previous section we call this equation the \textit{global} Gibbs-Duhem equation and we generalize it in the three mentioned ways.
To do so, we put $W=Y=0$ and $T=\delta(\phi-\phi')$ in the integrable part of the charge variation \eqref{integrable-charge-Y-VCT}, then we reach a local equation 
\begin{equation}\label{LEGD-eq-VCT}
\boxed{\bc{\bar{E}}=T_{_{\mathcal{N}}} \bc{\bar{S}} +{\cal U} \bc{\bar{J}}}
\end{equation}
which we interpret it as the \textit{local} Gibbs-Duhem equation. Similar to the local first law, this equation is also a local equation and its integrated version reduces to the usual Gibbs-Duhem relation for stationary cases \eqref{global GD}. It is important to note the thermodynamic variables which appear in this equation are subject to the equations of motion \eqref{EOM-VCT-1}.
\subsection{Local Zeroth Law} \label{sec:Non-expanding-no-news}
The goal of this subsection is to present a local expression for the zeroth law. Basically, we can think about the zeroth law as a statement of thermal equilibrium. In the usual thermodynamics, the flux of charges is proportional to the gradient of the chemical potentials, so we can take the absence of these kinds of flux as a statement of the zeroth law. So, these kinds of interpretations motivate us to put 
the flux part of the charge variation equal to zero. But we are going to take a weaker condition than what was mentioned. To do so, we start from the on-shell variation of the action
\begin{equation}\label{Onshell-deltaS-generic}
\begin{split}
    \delta I|_{\text{\tiny on-shell}}=\frac{1}{16\pi G} \int_{\mathcal{N}} \d v \d \phi \left[ -\bar{\mathcal{J}}\delta\mathcal{U}+\bar{\Omega}\delta \Gamma+\mathcal{D}_{v}\bar{\Omega}\,\delta\mathcal{P} \right]\, .
    \end{split}
\end{equation}
Here we have added $W$-term \eqref{W-Y-term-VCT} into the action. As a statement of local zeroth law, we require
\begin{equation}\label{zeroth-law-VCT-1}
    \delta I|_{\text{\tiny on-shell}}=\frac{1}{16\pi G} \int_{\mathcal{N}} \d v \d \phi \, \delta \mathcal{G}\, ,
\end{equation}
here we can interpret $\mathcal{G}$ as a boundary Lagrangian. This requirement \eqref{zeroth-law-VCT-1} results
\begin{equation}
    \delta\bc{\mathcal{G}}=-\bc{\bar{\mathcal{J}}}\delta\mathcal{U}-\bc{\bar{\mathcal{S}}}\delta T_{_\mathcal{N}}+\mathcal{D}_{v}\bc{\bar{\mathcal{S}}}\,\delta\bc{\mathcal{P}}\, .
\end{equation}
where $\bc{\mathcal{G}}=\mathcal{G}/16\pi G$. From the combination of the local first law \eqref{local-first-law-VCT} and the local zeroth law \eqref{zeroth-law-VCT-1}, we reach
\begin{equation}\label{Hamiltonian-VCT}
    \delta\bc{\bar{\mathcal{H}}}=\mathcal{U}\delta\bc{\bar{\mathcal{J}}}+T_{_\mathcal{N}}\delta\bc{\bar{\mathcal{S}}}+\mathcal{D}_{v}\bc{\bar{\mathcal{S}}}\,\delta\bc{\mathcal{P}}\, , \hspace{1 cm} \bc{\bar{\mathcal{H}}}=\bc{\bar{\mathcal{G}}}+T_{_\mathcal{N}}\bc{\bar{\mathcal{S}}}+\mathcal{U}\bc{\bar{\mathcal{J}}}\, .
\end{equation}
The integrability condition of this equation, $\delta( \delta\bc{\bar{\mathcal{H}}})=0$, leads to
\begin{equation}
    T_{_\mathcal{N}}=\frac{\delta\bc{\bar{\mathcal{H}}}}{\delta\bc{\bar{\mathcal{S}}}}\approx -\mathcal{D}_{v}\bc{\mathcal{P}}, \hspace{1 cm} \mathcal{D}_{v}{\bc{\bar{\mathcal{S}}}}=\frac{\delta\bc{\bar{\mathcal{H}}}}{\delta\bc{\mathcal{P}}}, \hspace{1 cm} \mathcal{U}=\frac{\delta\bc{\bar{\mathcal{H}}}}{\delta\bc{\bar{\mathcal{J}}}}\, .
\end{equation}
The first two equations are the Hamilton equations. It simply shows $\bc{\bar{\mathcal{H}}}$ plays the role of the Hamiltonian in the boundary system and $\{\bc{\bar{\mathcal{S}}}, \bc{\mathcal{P}}\}$ are Heisenberg conjugate and satisfy a Heisenberg algebra. It should be stressed that this boundary Hamiltonian \eqref{Hamiltonian-VCT} does not determine from our analysis, because it involves an arbitrary function $\bc{\mathcal{G}}$.\footnote{Determination of $\mathcal{G}$ is equivalent to the choice of boundary conditions (thermodynamic ensemble).} A careful analysis of Raychaudhuri and Damour equations and also the above integrability conditions lead to the following algebra
\begin{equation}\label{Charge-brackets-VCT}
\begin{split}
  &\{\bc{\bar{S}}(v,\phi), \bc{P}(v,\phi')\}=\delta(\phi-\phi'),\quad  \{\bc{\bar{S}}(v,\phi), \bc{\bar{S}}(v,\phi')\}= \{\bc{P}(v,\phi), \bc{P}(v,\phi')\}=0,\\
&\{\bc{\bar{S}}(v,\phi), \bc{\bar{J}}(v,\phi')\}= \bc{\bar{S}}(v,\phi'){\partial_{\phi}}\delta(\phi-\phi'), \\
&\{\bc{P}(v,\phi), \bc{\bar{J}}(v,\phi')\}= \left(\bc{P}(v,\phi'){\partial_{\phi}}+ \bc{P}(v,\phi){\partial_{\phi'}}+\frac{1}{2\pi}\partial_{\phi'}\right)\delta(\phi-\phi'), \\
 &\{\bc{\bar{J}}(v,\phi), \bc{\bar{J}}(v,\phi')\}=\left(\bc{\bar{J}}(v,\phi')\partial_{\phi}-\bc{\bar{J}}(v,\phi)\partial_{\phi'}+\frac{1}{8\pi G\, \mu}\partial_{\phi'}^{3}\right)\delta(\phi-\phi')\, .
\end{split}
\end{equation}
This is a Heisenberg  $\, \inplus \,$ Witt algebra. It has been shown \cite{Adami:2021sko} that there is a direct sum slicing in which the charge algebra becomes Heisenberg $\oplus$ Virasoro. In summary, we can take the existence of an algebra among the surface charges as a statement of the local zeroth law.
\paragraph{Integrable slicing.}
In this part, we write the charge variation in another slicing which yields an integrable expression for the charge variation: \textit{integrable slicing} \cite{Adami:2020amw,Adami:2020ugu,Adami:2021nnf,Adami:2021sko,Adami:2022ktn,Taghiloo:2022kmh}. Let us look at the following field dependent combination of the symmetry generators
\begin{equation}
    \hat{W}=W-\Gamma T\, \hspace{1 cm} \hat{Y}=Y+\mathcal{U}T\, , \hspace{1 cm}  \hat{T}=\mathcal{D}_{v}\bar{\Omega} T\, .
\end{equation}
In terms of these generators, we get 
\begin{equation}\label{charge-variation-Y-1}
    \begin{split}
        {\delta} Q(\xi) \approx \frac{1}{16 \pi G} \int_0^{2 \pi} \d \phi (& \hat{W}\delta\bar{\Omega}+\hat{Y}\delta\bar{\mathcal{J}}+\hat{T}\delta\mathcal{P} )\, .
    \end{split}
\end{equation}
Now, if we assume our new generators are field independent, $\delta\hat{T}=\delta\hat{Y}=\delta\hat{W}=0$, then we will find the following integrable expression
\begin{equation}\label{charge-variation-Y-1}
    \begin{split}
        Q(\xi) \approx \frac{1}{16 \pi G} \int_0^{2 \pi} \d \phi (& \hat{W}\bar{\Omega}+\hat{Y}\bar{\mathcal{J}}+\hat{T}\mathcal{P} ).
    \end{split}
\end{equation}
In this slicing, the symmetry algebra yields 
\begin{subequations}\label{Wb12-Tb12-Yb12}
\begin{align}
    &\hat{T}_{12}=\partial_{\phi}(\hat{Y}_{1}\hat{T}_{2})-(1\leftrightarrow 2),\\
    &\hat{W}_{12}=\hat{Y}_{1}\partial_{\phi}\hat{W}_{2}-(1\leftrightarrow 2),\\
    &\hat{Y}_{12}=\hat{Y}_{1}\partial_{\phi}\hat{Y}_{2}-(1\leftrightarrow 2).
\end{align}
\end{subequations}
An interesting property of this integrable slicing is that its structure constants are $v$-independent. The charge algebra is the same as the symmetry algebra with the following central terms
\begin{equation}\label{central-term-VCT}
    K_{\xi_1,\xi_2}=\frac{1}{16\pi G}\int_{0}^{2\pi} \d {} \phi \left[\hat{T}_{2}\hat{W}_{1}+2\hat{Y}_{2}\partial_{\phi}\hat{T}_{1}+\frac{1}{\mu}\hat{Y}_{2}\partial_{\phi}^{3}\hat{Y}_{1}-(1\leftrightarrow 2)\right]\, .
\end{equation}
Surprisingly, this algebra is the same as \eqref{Charge-brackets-VCT}. So, we learn the zeroth law brings us to integrable slicings \cite{Adami:2021kvx}.
\section{Null Surface Thermodynamic, NVCT Case} \label{sec: Null surface thermodynamics (NVCT)}
In this section we consider the null boundary thermodynamic for the full NVCT solution phase space. In comparison with the previous section, we turn on the chiral massive news in the solution phase space of the theory. As we will see, this bulk mode through interactions with boundary d.o.f takes our boundary thermodynamics out of thermal equilibrium. In this section, we present a local first law, local Gibbs-Duhem equation, and local zeroth law in presence of this hard mode. In the whole of this section, the on-shell equality $\approx$ means we apply the NVCT equations of motion
\begin{equation}\label{EOM-NVCT-1}
    \mathcal{E}_{ll}=\mathcal{E}_{l\phi}=\mathcal{E}=0\, .
\end{equation}
The following quantities will play key roles in our null boundary thermodynamic description for the NVCT case
\begin{equation}\label{definitions-J-N}
    \bar{\mathcal{J}}_{{\text{\tiny{NVCT}}}}:=\bar{\mathcal{J}}-\frac{2}{\mu}\left[\mathcal{T}_{\phi\phi}+\Omega \partial_{\phi}\left(\frac{\mathcal{T}_{l\phi}}{\Omega\Theta_{l}}\right)\right]\, , \hspace{1 cm}     {N}:=\frac{2\mathcal{T}_{ll}}{\Omega\Theta_{l}}\left[2\Omega-\bar{\Omega}+\frac{\mathcal{T}_{l\phi}}{\mu\Theta_{l}}\right]\, .
\end{equation}
From now on the index NVCT for different quantities indicates they take corrections w.r.t the VCT case. In other words, these quantities reduce to the corresponding quantities in the VCT case for $\mathcal{T}_{ll}=\mathcal{T}_{l\phi}=\mathcal{T}_{\phi\phi}=0$. The news function, $N$, is a crucial part of our thermodynamic picture in NVCT case. It is proportional to $\mathcal{T}_{ll}$ and captures the information about the massive bulk propagating mode of TMG.
\paragraph{Equations of motion in terms of charges.} 
NVCT equations of motion \eqref{EOM-NVCT-1} in terms of thermodynamic variables yield
\begin{subequations}\label{charge-eom-NVCT-1}
    \begin{align}
       \mathcal{D}_{v}\bar{\Omega}=  &\Theta_{l}(2\Omega-\bar{\Omega})+\frac{1}{\mu}\left[\mathcal{T}_{l\phi}+\partial_{\phi}\left(\frac{\mathcal{T}_{ll}}{\Theta_{l}}\right)\right]\, ,\\
        \mathcal{D}_{v}\mathcal{P}= & \Gamma+\frac{2\mathcal{T}_{ll}}{\Theta_{l}}\, , \\
        \mathcal{D}_{v}\bar{\mathcal{J}}_{\text{\tiny{NVCT}}}\approx & -\bar{\Omega}\partial_{\phi}\Gamma-\mathcal{D}_{v}\bar{\Omega}\partial_{\phi}\mathcal{P}+N\partial_{\phi}\Omega-\partial_{\phi}\left[\Omega N+2\mathcal{D}_{v}\bar{\Omega}+\frac{2\partial_{\phi}^{2}\mathcal{U}}{\mu}\right] \, ,
    \end{align}
\end{subequations}
\paragraph{Symplectic potential.} 
A straightforward calculation yields the following expression for the Lee-Wald symplectic potential
\begin{equation}
    \begin{split}
        \Theta_{\text{\tiny{LW}}}^{r}=&-{\bar{\mathcal{J}}}_{{\text{\tiny{NVCT}}}}\,\delta\mathcal{U}+\bar{\Omega}\delta \Gamma+\mathcal{D}_{v}\bar{\Omega}\,\delta\mathcal{P}-N\delta\Omega-\partial_{\phi}\left(\frac{\mathcal{T}_{ll}\delta\mathcal{P}}{\mu\Theta_{l}}\right)+\delta\left(\frac{\mathcal{T}_{ll}\mathcal{T}_{l\phi}}{\mu\,\Theta_{l}^{2}}\right)\\
        &-\frac{1}{\mu}\partial_{v}\mathcal{A}_{\text{\tiny{NVCT}}} +\frac{1}{\mu}\delta \mathcal{B}_{{\text{\tiny{NVCT}}}}\, .
    \end{split}
\end{equation}
where we have dropped out the total derivative terms w.r.t $\phi$ coordinate and the second line only involves a total derivative term w.r.t $v$ coordinate and a total variation term. Their explicit form is given by
\begin{equation}
    \begin{split}
        \mathcal{A}_{\text{\tiny{NVCT}}}=&\mathcal{A}+2\delta \Omega \, \frac{\mathcal{T}_{l\phi}}{\Omega\Theta_{l}}\, ,\\
        \mathcal{B}_{{\text{\tiny{NVCT}}}}=&\mathcal{B}+\frac{\mathcal{T}_{ll}}{\Theta_{l}}\left[2\Omega-\frac{1}{\mu}\left(\frac{\mathcal{J}}{\Omega}+\frac{\partial_{\phi}\mathcal{P}}{2}+\frac{2\partial_{\phi}\Omega}{\Omega}-\frac{\mathcal{T}_{l\phi}}{\Theta_{l}}\right)\right]\, .
    \end{split}
\end{equation}
One can absorb $\mathcal{A}_{{\text{\tiny{NVCT}}}}$ and $\mathcal{B}_{{\text{\tiny{NVCT}}}}$ terms into the following appropriate $Y$ and $W$ terms
\begin{equation}\label{Y-W-term-NVCT}
    \begin{split}
     Y^{\mu \nu}_{\text{\tiny{NVCT}}}[\delta g ; g]=& -\frac{\sqrt{-g}}{16 \pi G \mu} \, \epsilon^{\mu \nu \lambda} B^{\text{\tiny{NVCT}}}_\lambda [\delta g ; g]\, , \hspace{1 cm} B^{\text{\tiny{NVCT}}}_\lambda=B_\lambda+{2\delta \Omega} \frac{l^\alpha \mathcal{T}_{\alpha \lambda}}{\Omega\Theta_{l}}\, ,\\
     W^{\mu}_{\text{\tiny{NVCT}}}[g]=&\frac{\sqrt{-g}\mathcal{B}_{\text{\tiny{NVCT}}}}{\mu\, \Omega}n^{\mu}\, .\\
    \end{split}
\end{equation}
Again similar to the VCT case, we add these freedoms \eqref{Y-W-term-NVCT} to any symplectic quantities and drop out their LW index. In \cite{Adami:2021sko}, these kinds of freedoms were used to obtain genuine slicings in the NVCT case.
\paragraph{Symplectic form.} 
The Lee-Wald symplectic form for the NVCT case after the addition of $Y$-term \eqref{Y-W-term-NVCT} leads to
\begin{equation}\label{presymplectic}
    \begin{split}
     \Omega
        =  \frac{1}{16 \pi G} \int_{\mathcal{N}} \d v \d{}\phi \,\left[\delta\mathcal{U}\wedge \delta\bar{\mathcal{J}}_{\text{\tiny{NVCT}}}-\delta\Gamma\wedge \delta\bar{\Omega}-\delta\mathcal{P}\wedge\delta(\mathcal{D}_{v}\bar{\Omega})+\delta \Omega\wedge\delta N\right].
    \end{split}
\end{equation}
Compared to the VCT case, we have an extra conjugate pair, $\Omega$ and $N$, associated with chiral massive news. We treat $N$ as a charge and $\Omega$ as its chemical potential. It is consistent with our previous criterion to distinguish chemical potentials and charges. Here $N$ carries information about the propagating mode and hence it depends on the theory. A comparison with the Einstein gravity in higher dimensions is also helpful. In \cite{Adami:2021nnf} it has been shown that the conjugate of news tensor is the metric on a co-dimension two cross section of the null boundary (transverse surface). In our three dimensional case, this transverse surface is a circle and $\Omega$ plays the role of its metric. In summary in the NVCT case, we have four chemical potentials, $\{\mathcal{U},\Gamma;\mathcal{P},\Omega\}$, and their four associated charges, $\{\bar{\mathcal{J}}_{\text{\tiny{NVCT}}},\bar{\Omega};\mathcal{D}_{v}\bar{\Omega},N\}$.
\paragraph{Charge variations.} 
The transformation laws for the labels of the boundary d.o.f are
\begin{subequations}\label{delta-charges-NVCT}
    \begin{align}      
         \delta_{\xi}\bar{\Omega}=& (T\mathcal{D}_{v}+\mathcal{L}_{\hat{Y}})\bar{\Omega}+\frac{\mathcal{T}_{ll}\partial_{\phi}T}{\mu\,\Theta_{l}}\, ,\\
        \delta_\xi {\cal P} = &(T\mathcal{D}_{v}+\mathcal{L}_{\hat{Y}})\mathcal{P}-W\, ,\label{delta-P}\\
        \delta_\xi (\mathcal{D}_{v}\bar{\Omega})= & \mathcal{D}_{v}(T\mathcal{D}_{v}\bar{\Omega})+\mathcal{L}_{\hat{Y}}(\mathcal{D}_{v}\bar{\Omega})+\mathcal{D}_{v}\left(\frac{\mathcal{T}_{ll}\partial_{\phi}T}{\mu\Theta_{l}}\right)\, ,\\
      \delta_{\xi}\bar{\mathcal{J}}_{\text{\tiny{NVCT}}}\approx & (T\mathcal{D}_{v}+\mathcal{L}_{\hat{Y}})\bar{\mathcal{J}}_{\text{\tiny{NVCT}}}+\bar{\Omega}\partial_{\phi}W-\left(2\mathcal{D}_{v}\bar{\Omega}+\bar{\Omega}\Gamma+\Omega N+\frac{\mathcal{T}_{ll}\partial_{\phi}\mathcal{P}}{\mu\, \Theta_{l}}\right)\partial_{\phi}T\nonumber\\
      &+\frac{2}{\mu}\left[T\partial_{\phi}^{3}\mathcal{U}-\partial_{\phi}^{3}\hat{Y}-\partial_{\phi}\left(\frac{\mathcal{T}_{ll}\partial_{\phi}T}{\Theta_{l}}\right)\right]\, ,
\end{align}
\end{subequations}
and the bulk d.o.f transforms under the large diffeomorphisms as
\begin{equation}
          \delta_{\xi}N=\mathcal{D}_{v}(T\,N)+\mathcal{L}_{\hat{Y}}N\, .
\end{equation}
The variation of $N$ has a nice property: it is homogeneous in $N$ \cite{Adami:2021nnf,Compere:2018ylh}. This means by large gauge transformations we can not create the bulk gravitons (hard propagating mode). This is the property that we expect to be true for any hard modes.
\subsection{Surface Charge Variation} 
The covariant phase space method \cite{Iyer:1994ys,Tachikawa:2006sz, Kim:2013cor} gives the following expression for the charge variation
\begin{equation}\label{charge-variation-Y-NVCT-1}
   \begin{split}
        \slashed{\delta} Q_{{\text{\tiny{NVCT}}}}(\xi) \approx \frac{1}{16 \pi G} \int_0^{2 \pi} \d \phi & \left(W \delta\bar{\Omega}+Y\delta \bar{\mathcal{J}}_{{\text{\tiny{NVCT}}}}+T \slashed{\delta}\bar{\mathcal{H}}_{{\text{\tiny{NVCT}}}}\right) \, ,
    \end{split}
\end{equation}
where
\begin{equation}
    \slashed{\delta}\bar{\mathcal{H}}_{{\text{\tiny{NVCT}}}}:=-\Gamma\delta\bar{\Omega}+\mathcal{U}\delta  \bar{\mathcal{J}}_{{\text{\tiny{NVCT}}}}+\mathcal{D}_{v}\bar{\Omega}\delta\mathcal{P}-{N}\delta\Omega-\partial_{\phi}\left(\frac{\mathcal{T}_{ll}\delta\mathcal{P}}{\mu\Theta_{l}}\right)\, .
\end{equation}
Here we have used $Y$ term \eqref{Y-W-term-NVCT} and on-shell conditions \eqref{EOM-NVCT-1} for the NVCT case.
One can split the charge variation into the integrable and flux parts as
\begin{equation}\label{integrable-charge-Y-NVCT}
    Q_{{\text{\tiny{NVCT}}}}^{\text{I}}(\xi)=\frac{1}{16 \pi G} \int_0^{2 \pi} \d \phi \left[ W \bar{\Omega}+Y  \bar{\mathcal{J}}_{{\text{\tiny{NVCT}}}}+T\left(\mathcal{U}{ \bar{\mathcal{J}}_{{\text{\tiny{NVCT}}}}}-\Gamma \bar{\Omega}-\frac{\mathcal{T}_{ll}\mathcal{T}_{l\phi}}{\mu\,\Theta_{l}^{2}}\right)\right]\, ,
\end{equation}
and
\begin{equation}\label{Non-Integrable-Charge-Y-NVCT}
    \mathcal{F}_{{\text{\tiny{NVCT}}}}(\xi)=\frac{1}{16 \pi G} \int_0^{2 \pi} \d \phi \, T \Biggl\{- \bar{\mathcal{J}}_{{\text{\tiny{NVCT}}}}\,\delta\mathcal{U}+\bar{\Omega}\,\delta \Gamma+\mathcal{D}_{v}\bar{\Omega}\,\delta\mathcal{P}-N\delta\Omega-\partial_{\phi}\left(\frac{\mathcal{T}_{ll}\delta\mathcal{P}}{\mu\Theta_{l}}\right)+\delta\left(\frac{\mathcal{T}_{ll}\mathcal{T}_{l\phi}}{\mu\,\Theta_{l}^{2}}\right)\Biggr\}\, .
\end{equation}
This splitting yields the same central charge as we had in the VCT case \eqref{central-charge-1}.
The zero mode charges for surface charge variation \eqref{charge-variation-Y-NVCT-1} result
\begin{equation}\label{Charges-zero-mode-full-NVCT}
\begin{split}
        Q_{{\text{\tiny{NVCT}}}}({-r\partial_r}) &:= \frac{\mathbf{\bar{S}}}{4\pi}=\frac{1}{16\pi G}  \int_0^{2 \pi} \d \phi\, \bar{\Omega}\, ,     \\ 
        Q_{{\text{\tiny{NVCT}}}}({\partial_{\phi}}) &:=\mathbf{\bar{J}}_{\text{\tiny{NVCT}}}=\frac{1}{16\pi G}  \int_0^{2 \pi} \d \phi\, \bar{\mathcal{J}}_{{\text{\tiny{NVCT}}}}\, ,\\
       \slashed{\delta} Q_{{\text{\tiny{NVCT}}}}({\partial_v}) &:= \slashed{\delta}\mathbf{\bar{H}}_{\text{\tiny{NVCT}}}=\frac{1}{16\pi G}  \int_0^{2 \pi} \d \phi\, \left(-\Gamma\delta\bar{\Omega}+\mathcal{U}\delta  \bar{\mathcal{J}}_{{\text{\tiny{NVCT}}}}+\mathcal{D}_{v}\bar{\Omega}\delta\mathcal{P}-{N}\delta\Omega \right) \, .
\end{split}
\end{equation}
We can also read zero mode charges for the integrable part of the charge variation \eqref{integrable-charge-Y-NVCT}
\begin{equation}\label{Charges-zero-mode-integrable-NVCT}
\begin{split}
        Q^{\text{I}}_{{\text{\tiny{NVCT}}}}({-r\partial_r}) &:= \frac{\mathbf{\bar{S}}}{4\pi}=\frac{1}{16\pi G}  \int_0^{2 \pi} \d \phi\, \bar{\Omega}\, ,     \\ 
        Q^{\text{I}}_{{\text{\tiny{NVCT}}}}({\partial_{\phi}}) &:= \mathbf{\bar{J}}_{\text{\tiny{NVCT}}}=\frac{1}{16\pi G}  \int_0^{2 \pi} \d \phi\, \bar{\mathcal{J}}_{{\text{\tiny{NVCT}}}}\, ,\\
       Q^{\text{I}}_{{\text{\tiny{NVCT}}}}({\partial_v}) &:= \mathbf{\bar{E}}_{\text{\tiny{NVCT}}}=\frac{1}{16\pi G}  \int_0^{2 \pi} \d \phi\, \left(\mathcal{U}{ \bar{\mathcal{J}}_{{\text{\tiny{NVCT}}}}}-\Gamma \bar{\Omega}-\frac{\mathcal{T}_{ll}\mathcal{T}_{l\phi}}{\mu\,\Theta_{l}^{2}}\right) \,.
\end{split}
\end{equation}
We observe the angular momentum and energy take corrections in comparison with the VCT case but the entropy aspect charge does not.
\paragraph{Balance equation.}
Similar to the VCT case we obtain the following balance equation among different parts of the charge variation \eqref{charge-variation-Y-NVCT-1}
\begin{equation}
{\frac{\d {}}{\d v} Q_{\xi}^{{\text{\tiny{I}}}}\approx-\mathcal{F}_{\partial_{v}}(\delta_{\xi}g;g)+{K}_{\xi,\partial_{v}}\, .}
\end{equation}
One can interpret this equation in several ways. The first way is to consider this equation as a manifestation of boundary equations of motion which have been written in terms of the surface charges. The second interpretation of this equation is that it states how our boundary d.o.f adjust themselves due to the passage of bulk chiral mode. In other words, this equation describes interactions between bulk and boundary d.o.f.
\subsection{Null Boundary Thermodynamical Phase Space, NVCT Case}\label{sec:Thermodynamical phase space-NVCT}
Here we present our general boundary thermodynamic picture for the full solution phase space of TMG.  
\begin{enumerate}
\item[I.] Null boundary solution space for the NVCT case consists of the following three parts: 
\begin{enumerate}
    \item[I.1)] thermodynamic sector: parametrized by $(\Gamma, {\cal U})$ and conjugate charges $(\bar{\Omega}, \bar{\mathcal{J}}_{\text{\tiny{NVCT}}})$. They are subject to equations of motion \eqref{charge-eom-NVCT-1}.
    \item[I.2)] ${\cal P}$, which only appears in  the flux \eqref{Non-Integrable-Charge-Y-NVCT} and not in the integrable charge \eqref{integrable-charge-Y-NVCT}. We observe again even in the presence of bulk mode, this quantity does not accept any corrections. We conjecture that it is a universal property of any expanding null surface in any general covariant theories of gravity. Similar to the VCT case, its conjugate is equal to the time derivative of the entropy aspect.
    \item[I.3)] The bulk mode is parameterized by $\Omega$ and its `conjugate charge' $N$ which appear in the flux \eqref{Non-Integrable-Charge-Y-NVCT}.
\end{enumerate}
    \item[II.] We have two sources that make our boundary system out of thermal equilibrium: bulk effect, $N$, through interactions with boundary d.o.f, and boundary effect which is parametrized by $\mathcal{P}$.
    \item[III.] The time derivatives of entropy aspect $\mathcal{D}_{v}\bar{\Omega}$ and area density $\mathcal{D}_{v}{\Omega}$  measure the OTE from the bulk and boundary viewpoints respectively.
\end{enumerate}
\subsection{Local First Law at Null Boundary}\label{subsec:first-law}
By doing the same procedure as we have done in the VCT case, we reach the following equation for the local first law
\begin{equation}\label{local-first-law}
\boxed{\slashed{\delta} \bc{\bar{H}}_{{\text{\tiny{NVCT}}}}=T_{_{\mathcal{N}}} \delta\bc{\bar{S}} +{\cal U}\delta\bc{\bar{{J}}}_{{\text{\tiny{NVCT}}}} +\mathcal{D}_{v}\bc{\bar{S}}\,\delta \bc{P}-\bc{N}\delta\bc{{S}}-\partial_{\phi}\left(\frac{\bc{{T}}_{ll}\delta\bc{P}}{\mu\Theta_{l}}\right)}
\end{equation}
where 
\begin{equation}\label{local-first-law-NVCT-1}
    \bc{{\mathcal{S}}}=\frac{{\Omega}}{4 \pi}\, ,\hspace{1 cm} \bc{N}=\frac{N}{4 G}\, ,  \hspace{1 cm} T_{_{\mathcal{N}}}:= -\frac{\Gamma}{4\pi}\, , \hspace{1 cm} \bc{\bar{{J}}}_{{\text{\tiny{NVCT}}}}=\frac{{\bar{\mathcal{J}}}_{{\text{\tiny{NVCT}}}}}{16\pi G}  \,, \hspace{1 cm} \bc{{T}}_{ll}=\frac{{{T}}_{ll}}{4 G}\, .
\end{equation}
Once again, this is a local equation in boundary coordinates. In comparison with the VCT case, we have two further terms. One of them which is parametrized by $N$ arises due to the existence of chiral massive news in the bulk. This is a bulk effect which through interactions with boundary d.o.f takes the boundary thermodynamic out of thermal equilibrium. The last term in \eqref{local-first-law-NVCT-1} is proportional to $1/\mu$ and hence is a property of TMG. This term appears as a total derivative term w.r.t $\phi$ coordinate and is reminiscent of the fact that our equation is local. It is a combination of bulk and boundary effects (involves both $\mathcal{P}$ and $\mathcal{T}_{ll}$).
\subsection{Local Gibbs-Duhem Equation at Null Boundary}\label{subsec:GD-relation}
The local Gibbs-Duhem equation for the NVCT case is
\begin{equation}\label{LEGD-eq-NVCT}
\boxed{\bc{\bar{E}}_{{\text{\tiny{NVCT}}}}=T_{_{\mathcal{N}}} \bc{\bar{S}} +{\cal U} \bc{\bar{{J}}}_{{\text{\tiny{NVCT}}}}-\frac{1}{\mu}\frac{\bc{\mathcal{T}}_{ll}\bc{\mathcal{T}}_{l\phi}}{\Theta_{l}^{2}}.}
\end{equation}
where $\bc{\mathcal{T}}_{l\phi}=\frac{{\mathcal{T}}_{l\phi}}{4\pi}$. In comparison with the VCT case, it has an extra term, which is proportional to $\mathcal{T}_{ll}$ and parameterizes the bulk effect of massive chiral news. This extra term does not have any counterpart in $D$-dimensional Einstein gravity \cite{Adami:2021kvx} so it is a property of the underlying theory.
\subsection{Local Zeroth Law} \label{sec:Non-expanding-no-news}
In this section, we discuss the zeroth law for the NVCT case. We start with the on-shell variation of the action
\begin{equation}\label{Onshell-deltaS-generic}
\begin{split}
    \hspace{-.1 cm}\delta I|_{\text{\tiny on-shell}}=\frac{1}{16\pi G} \int_{\mathcal{N}} \d v \d \phi \, \Biggl\{ -{\bar{\mathcal{J}}}_{{\text{\tiny{NVCT}}}}\,\delta\mathcal{U}+\bar{\Omega}\delta \Gamma+\mathcal{D}_{v}\bar{\Omega}\,\delta\mathcal{P}-N\delta\Omega-\partial_{\phi}\left(\frac{\mathcal{T}_{ll}\delta\mathcal{P}}{\mu\Theta_{l}}\right)+\delta\left(\frac{\mathcal{T}_{ll}\mathcal{T}_{l\phi}}{\mu\,\Theta_{l}^{2}}\right)\Biggr\}\, .
    \end{split}
\end{equation}
Similar to the VCT case, we take the following requirement as a statement of the local zeroth law 
\begin{equation}
    \delta I|_{\text{\tiny on-shell}}=\frac{1}{16\pi G} \int_{\mathcal{N}} \d v \d \phi \, \delta \mathcal{G}_{{\text{\tiny{NVCT}}}}\, ,
\end{equation}
hence, we get
\begin{equation}\label{local-zeroth-law-NVCT-1}
    \delta\bc{\mathcal{G}}_{{\text{\tiny{NVCT}}}}=-\bc{\bar{{J}}}_{{\text{\tiny{NVCT}}}}\delta\mathcal{U}-\bc{\bar{\mathcal{S}}}\delta T_{_\mathcal{N}}+\mathcal{D}_{v}\bc{\bar{\mathcal{S}}}\,\delta\bc{\mathcal{P}}-\bc{N}\delta\bc{S}-\partial_{\phi}\left(\frac{\bc{\mathcal{T}}_{ll}\delta\bc{\mathcal{P}}}{\mu\Theta_{l}}\right)+\delta\left(\frac{\bc{\mathcal{T}}_{ll}\bc{\mathcal{T}}_{l\phi}}{\mu\,\Theta_{l}^{2}}\right)\, .
\end{equation}
The integrability condition, $\delta(\delta \bc{\mathcal{G}}_{{\text{\tiny{NVCT}}}})=0$, leads to an equation such as $\sum_{\alpha, \beta} C_{\alpha\beta}\delta Q_{\alpha}\wedge \delta Q_{\beta}\approx 0$, where $Q_{\alpha}$ denote generic charges and $C_{\alpha\beta}$ is a skew-symmetric matrix. To get this result, we have used the equations of motion \eqref{charge-eom-NVCT-1}. In other words, we need to substitute chemical potentials in terms of the charges. Because our charges are independent so the unique solution of this equation is $C_{\alpha\beta}=0$. An immediate consequence of this equation is  $\mathcal{T}_{ll}=0$. The vanishing of $\mathcal{T}_{ll}$ leads to the vanishing of the news function, $N=0$. Therefore we observe the zeroth law enforces us to turn off the bulk mode \cite{Adami:2021kvx}. Putting these conditions $\mathcal{T}_{ll}=0$ and $N=0$, into the zeroth law \eqref{local-zeroth-law-NVCT-1} and combining it with the local first law \eqref{local-first-law-NVCT-1} result
\begin{equation}\label{local-zeroth-law-NVCT-2}
    \slashed{\delta}\bc{\bar{\mathcal{H}}}_{{\text{\tiny{NVCT}}}}=\mathcal{U}\delta\bc{\bar{{J}}}_{{\text{\tiny{NVCT}}}}+T_{_\mathcal{N}}\delta\bc{\bar{\mathcal{S}}}+\mathcal{D}_{v}\bc{\bar{\mathcal{S}}}\,\delta\bc{\mathcal{P}}\, , \hspace{1 cm}  \bc{\bar{\mathcal{H}}}_{{\text{\tiny{NVCT}}}}:=\bc{\bar{\mathcal{G}}}+T_{_\mathcal{N}}\bc{\bar{\mathcal{S}}}+\mathcal{U}\bc{\bar{{J}}}_{{\text{\tiny{NVCT}}}}\, .
\end{equation}
The remaining integrability conditions to have nontrivial solutions for \eqref{local-zeroth-law-NVCT-2}, yield
\begin{equation}
    T_{_\mathcal{N}}=-\frac{\delta\bc{\bar{\mathcal{H}}_{{\text{\tiny{NVCT}}}}}}{\delta\bc{\bar{\mathcal{S}}}}, \hspace{1 cm} \mathcal{D}_{v}{\bc{\bar{\mathcal{S}}}}=\frac{\delta\bc{\bar{\mathcal{H}}_{{\text{\tiny{NVCT}}}}}}{\delta\bc{\mathcal{P}}}, \hspace{1 cm} \mathcal{U}=\frac{\delta\bc{\bar{\mathcal{H}}_{{\text{\tiny{NVCT}}}}}}{\delta\bc{\bar{{J}}}_{{\text{\tiny{NVCT}}}}}\, .
\end{equation}
Again these equations enforce an algebra among the charges. This algebra is the same as the VCT case \eqref{Charge-brackets-VCT}.
\section{Thermodynamics of Non-Expanding Null Surfaces} \label{Non-expanding case}
In all previous calculations, we assumed $\Theta_{l}\neq 0$. In this section, we consider a special sector of the solution phase space, the \textit{non-expanding} case ($\Theta_{l}=0$). From the definition of $\mathcal{T}_{ll}$ \eqref{Tmunu-components} for the non-expanding case, we find $\mathcal{T}_{ll}=0$. Because the news tensor is proportional with $\mathcal{T}_{ll}$ \eqref{definitions-J-N} so in this case, we do not have any bulk propagating mode, $N=0$. In this section, we will show there is also a local thermodynamic description for non-expanding null surfaces. Similar to the expanding null surfaces, we will present local thermodynamic equations for this case. It is important to point out that this case is not corresponding to the VCT case, because even though $\mathcal{T}_{ll}=0$, we still have $\mathcal{T}_{l\phi}\neq 0$ and $\mathcal{T}_{\phi\phi}\neq 0$.
\paragraph{Symplectic potential.} 
In this case the Lee-Wald symplectic potential up to total derivative terms w.r.t $\phi$, yields
\begin{equation}
  16\pi G\,  \Theta_{\text{\tiny{NE,LW}}}^{r}=-\bar{\mathcal{J}}_{\text{\tiny{NE}}}\delta\mathcal{U}+\bar{\Omega}_{\text{\tiny{NE}}}\delta \Gamma+\frac{1}{\mu}\mathcal{T}_{l\phi}\delta\mathcal{P}_{\text{\tiny{NE}}}-\frac{1}{\mu}\partial_{v}\mathcal{A}_{\text{\tiny{NE}}} +\frac{1}{\mu}\delta \mathcal{B}_{\text{\tiny{NE}}}
\end{equation}
where the thermodynamic quantities for non-expanding case are
\begin{equation}
    \begin{split}
    \bar{\Omega}_{\text{\tiny{NE}}}=&\Omega+\frac{1}{\mu}\left[\frac{\mathcal{J}}{2\Omega}+\frac{2\partial_{\phi}\Omega}{\Omega}+\frac{1}{2}\partial_{\phi}\mathcal{P}_{\text{\tiny{NE}}}\right]\, , \hspace{1 cm} \mathcal{P}_{\text{\tiny{NE}}}=\ln\left(\frac{\eta}{\Omega^{2}\mathcal{T}_{l\phi}^2}\right)\, ,\\
         \bar{\mathcal{J}}_{\text{\tiny{NE}}}=&\mathcal{J}+\frac{1}{\mu}\left[\left(\frac{\mathcal{J}}{2\Omega}\right)^2+\left(\frac{\Omega}{l}\right)^2-2\mathcal{T}_{\phi\phi}-\left(\frac{\partial_{\phi}\mathcal{P}_{\text{\tiny{NE}}}}{2}\right)^2-\frac{2\partial_{\phi}\Omega\partial_{\phi}\mathcal{P}_{\text{\tiny{NE}}}}{\Omega}-8\partial_{\phi}\left(\frac{\partial_{\phi}\Omega}{\Omega}\right)\right] \, .
    \end{split}
\end{equation}
The total derivative term and total variation term are
\begin{equation}
    \begin{split}
        \mathcal{A}_{\text{\tiny{NE}}}=&\frac{\delta \Omega \partial_\phi \Omega}{2 \Omega^2}+\frac{\delta \eta \partial_\phi \eta}{4 \eta^2}+ \Omega^2 \Theta_{n} \, \delta \mathcal{U} + \delta \omega+\frac{2\delta\Omega}{\Omega}\partial_{\phi}\mathcal{P}_{{\text{\tiny{NE}}}}+\frac{1}{4}\delta\mathcal{P}_{{\text{\tiny{NE}}}} \partial_{\phi}\mathcal{P}_{{\text{\tiny{NE}}}} \, ,\\
        \mathcal{B}_{\text{\tiny{NE}}}=&\frac{1}{2}\Omega^{2}\Theta_{n}\mathcal{D}_{v}\mathcal{U}+\partial_{v}\omega-{\partial_{\phi}\mathcal{U}}\left(\frac{1}{2}\mathcal{U}\Omega^2\Theta_{n}+\frac{\mathcal{J}}{4\Omega}-\frac{8\partial_{\phi}\Omega}{\Omega}+\frac{\partial_{\phi}\eta}{4\eta}-\partial_{\phi}\mathcal{P}_{\text{\tiny{NE}}}\right)\\
        &-\left(\frac{\mathcal{J}}{4\Omega}+\frac{\partial_{\phi}\mathcal{P}_{\text{\tiny{NE}}}}{4}+\frac{2\partial_{\phi}\Omega}{\Omega}\right)\Gamma\, .
    \end{split}
\end{equation}
We introduce the following $W$ and $Y$ terms to remove these terms
\begin{equation}\label{W-Y-term-NE}
    \begin{split}
        W_{\text{\tiny{NE}}}^{\mu}[g]=\frac{\sqrt{-g}\mathcal{B}_{\text{\tiny{NE}}}}{\mu\, \Omega}n^{\mu}\, , \hspace{1 cm}    Y_{\text{\tiny{NE}}}^{\mu \nu}[\delta g ; g]= -\frac{\sqrt{-g}}{16 \pi G \mu} \, \epsilon^{\mu \nu \lambda} B^{{\text{\tiny{NE}}}}_\lambda [\delta g ; g]\, ,
    \end{split}
\end{equation}
with
\begin{equation}
    B^{{\text{\tiny{NE}}}}_\lambda [\delta g ; g] = \frac{1}{4} \,\Gamma^\alpha_{\lambda \beta} \delta g^\beta_\alpha -\, n_\alpha l^\beta \delta \Gamma^{\alpha}_{\lambda \beta}{+ \frac{2\delta \Omega}{ \Omega} \partial_\lambda \mathcal{P}_{{\text{\tiny{NE}}}}}+\frac{1}{4}\delta\mathcal{P}_{{\text{\tiny{NE}}}}\partial_{\lambda}\mathcal{P}_{{\text{\tiny{NE}}}} \, .
\end{equation}
By using these $Y$ and $W$ terms, we reach
\begin{equation}
  16\pi G\,  \Theta_{\text{\tiny{NE}}}^{r}= -\bar{\mathcal{J}}_{\text{\tiny{NE}}}\delta\mathcal{U}+\bar{\Omega}_{\text{\tiny{NE}}}\delta \Gamma+\frac{1}{\mu}\mathcal{T}_{l\phi}\delta\mathcal{P}_{\text{\tiny{NE}}}\, .
\end{equation}
We add these freedoms \eqref{W-Y-term-NE} to any Lee-Wald quantities in the non-expanding case and drop their LW index. Again these kinds of $Y$-terms lead to the integrable expression for the surface charges in the non-expanding case \cite{Adami:2021sko}.
\paragraph{Equations of motion in terms of charges.} 
In this case, the Einstein equations lead to the following simple equations
\begin{equation}\label{EOM-NE-1}
    \mathcal{D}_{v}\bar{\mathcal{J}}_{\text{\tiny{NE}}}=-\mathcal{D}_{v}\bar{\Omega}_{\text{\tiny{NE}}}\partial_{\phi}\mathcal{P}_{\text{\tiny{NE}}}-\bar{\Omega}_{\text{\tiny{NE}}}\partial_{\phi}\Gamma-\frac{4}{\mu}\partial_{\phi}\left(\mathcal{T}_{l\phi}+2\partial_{\phi}^{2}\mathcal{U}\right)\, , \hspace{1cm} \mathcal{D}_{v}\bar{\Omega}_{\text{\tiny{NE}}}=\frac{\mathcal{T}_{l\phi}}{\mu}, \hspace{1 cm}  \mathcal{D}_{v}\mathcal{P}_{\text{\tiny{NE}}}=\Gamma\, .
\end{equation} 
\paragraph{Symplectic form.} 
The symplectic form for the non-expanding case is given by
\begin{equation}\label{presymplectic-NE}
    \begin{split}
     \Omega
        \approx  \frac{1}{16 \pi G} \int_{\mathcal{N}} \d v \d{}\phi \,\left[\delta\mathcal{U}\wedge \delta\bar{\mathcal{J}}_{\text{\tiny{NE}}}-\delta\Gamma\wedge \delta\bar{\Omega}_{\text{\tiny{NE}}}-\delta\mathcal{P}_{\text{\tiny{NE}}}\wedge\delta(\mathcal{D}_{v}\bar{\Omega}_{\text{\tiny{NE}}})\right]
    \end{split}
\end{equation}
here we have used \eqref{EOM-NE-1}. We recognize three thermodynamic conjugate pairs: there are three chemical potentials $\{\Gamma, \mathcal{U} , \mathcal{P}_{\text{\tiny{NE}}}\}$ and their corresponding thermodynamic charges $\{\bar{\Omega}_{\text{\tiny{NE}}},\bar{\mathcal{J}}_{\text{\tiny{NE}}},\mathcal{D}_{v}\bar{\Omega}_{\text{\tiny{NE}}}\}$. The off-shell chemical potential associated with $\mathcal{P}_{\text{\tiny{NE}}}$ is equal to $\mathcal{T}_{l\phi}/\mu$. From the definition of $\mathcal{T}_{l\phi}$ \eqref{Tmunu-components} for the non-expanding case, one can interpret this quantity geometrically as a time derivative of the twist field. So, it seems for the non-expanding case, the twist field plays the role of the expansion. On the other hand, the on-shell chemical potential of  $\mathcal{P}_{\text{\tiny{NE}}}$ is given by the time derivative of the Wald entropy aspect. It is worth emphasizing that the last chemical potential, $\mathcal{P}_{\text{\tiny{NE}}}$, does not have any Einstein counterpart.
\paragraph{Charge variations.} 
The variation of the surface charges is given by\footnote{It should be emphasized that ${\mathcal{P}}_{\text{\tiny{NE}}}$ does not have a well-defined weight. We define its derivative as
\begin{equation}
    \mathcal{D}_{v}{\mathcal{P}}_{\text{\tiny{NE}}}:=\partial_{v}{\mathcal{P}}_{\text{\tiny{NE}}}-\mathcal{L}_{\mathcal{U}}\mathcal{P}_{\text{\tiny{NE}}}, \hspace{1 cm}  \mathcal{L}_{\mathcal{U}}{\mathcal{P}}_{\text{\tiny{NE}}}:=\mathcal{U}\partial_{\phi}{\mathcal{P}}_{\text{\tiny{NE}}}-4\partial_{\phi}\mathcal{U}\, .
\end{equation}}
\begin{equation}
\begin{split}
    \delta_{\xi}\bar{\Omega}_{\text{\tiny{NE}}}=&(T\mathcal{D}_{v}+\mathcal{L}_{\hat{Y}})\bar{\Omega}_{\text{\tiny{NE}}}\, ,\\
    \delta_{\xi}{\mathcal{P}}_{\text{\tiny{NE}}}=& (T\mathcal{D}_{v}+\mathcal{L}_{\hat{Y}}){\mathcal{P}}_{\text{\tiny{NE}}}-W\, ,\\
    \delta_\xi (\mathcal{D}_{v}\bar{\Omega}_{\text{\tiny{NE}}})= & \mathcal{D}_{v}(T\mathcal{D}_{v}\bar{\Omega}_{\text{\tiny{NE}}})+\mathcal{L}_{\hat{Y}}(\mathcal{D}_{v}\bar{\Omega}_{\text{\tiny{NE}}})\, ,\\
    \delta_{\xi}\bar{\mathcal{J}}_{\text{\tiny{NE}}}=&
    (T\mathcal{D}_{v}+\mathcal{L}_{\hat{Y}}){\bar{\mathcal{J}}_{\text{\tiny{NE}}}}+\bar{\Omega}_{\text{\tiny{NE}}}(\partial_{\phi}W-\Gamma\partial_{\phi}T)+\frac{8}{\mu}\left(-\frac{\mathcal{T}_{l\phi}}{2}\partial_{\phi}T+T\partial_{\phi}^{3}\mathcal{U}-\partial_{\phi}^{3}\hat{Y}\right)\, .
\end{split}
\end{equation}
\subsection{Surface Charge Variation} 
In this case for the surface charge variation, we get
\begin{equation}\label{charge-variation-NE-1}
\begin{split}
    \slashed{\delta} Q_{\text{\tiny{NE}}}(\xi) \approx  \frac{1}{16 \pi G}   \int_0^{2 \pi} \d \phi \, \biggl\{  W\delta\bar{\Omega}_{\text{\tiny{NE}}}+Y\delta\bar{\mathcal{J}}_{\text{\tiny{NE}}}+T \left(\mathcal{U}\delta\bar{\mathcal{J}}_{\text{\tiny{NE}}}-\Gamma\delta\bar{\Omega}_{\text{\tiny{NE}}}+\frac{1}{\mu}\mathcal{T}_{l\phi}\delta\mathcal{P}_{\text{\tiny{NE}}} \right)\biggr\}\, .
    \end{split}
\end{equation}
Similar to the previous cases, one can separate the charge variation into the integrable and non-integrable parts 
\begin{equation}\label{integrable-Non-Integrable-charge-Y-NE}
    \begin{split}
        Q_{\text{\tiny{NE}}}^{\text{I}}(\xi)=&\frac{1}{16 \pi G} \int_0^{2 \pi} \d \phi \left[ W \bar{\Omega}_{\text{\tiny{NE}}}+Y \bar{\mathcal{J}}_{\text{\tiny{NE}}}+T(\mathcal{U}\bar{\mathcal{J}}_{\text{\tiny{NE}}}-\Gamma \bar{\Omega}_{\text{\tiny{NE}}})\right]\, ,\\
        \mathcal{F}_{\text{\tiny{NE}}}({\xi})=&\frac{1}{16 \pi G} \int_0^{2 \pi} \d \phi \, T \left(-\bar{\mathcal{J}}_{\text{\tiny{NE}}}\delta\mathcal{U}+\bar{\Omega}_{\text{\tiny{NE}}}\delta \Gamma+\frac{1}{\mu}\mathcal{T}_{l\phi}\delta\mathcal{P}_{\text{\tiny{NE}}}\right)\, .
    \end{split}
\end{equation}
This leads to the following central term
\begin{equation}
    K^{\text{\tiny{NE}}}_{\xi_1,\xi_2}=\frac{1}{4\pi G\, \mu}\int_{0}^{2\pi} \d \phi (\hat{Y}_{2}\partial_{\phi}^{3}\hat{Y}_{1}-\hat{Y}_{1}\partial_{\phi}^{3}\hat{Y}_{2})\, .
\end{equation}
This central term up to a numerical factor 4 is matched with \eqref{central-charge-1}.
In the non-expanding case, one can also read the zero mode charges from the full form of charge variation \eqref{charge-variation-Y-1-NE}
\begin{equation}\label{Charges-zero-mode-full-NE}
\begin{split}
        Q_{{\text{\tiny{NE}}}}({-r\partial_r}) &:= \frac{\mathbf{\bar{S}}_{{\text{\tiny{NE}}}}}{4\pi}=\frac{1}{16\pi G}  \int_0^{2 \pi} \d \phi\, \bar{\Omega}_{{\text{\tiny{NE}}}}\, ,     \\ 
        Q_{{\text{\tiny{NE}}}}({\partial_{\phi}}) &:= \mathbf{\bar{J}}_{\text{\tiny{NE}}}=\frac{1}{16\pi G}  \int_0^{2 \pi} \d \phi\, \bar{\mathcal{J}}_{{\text{\tiny{NE}}}}\, ,\\
       \slashed{\delta} Q_{{\text{\tiny{NE}}}}({\partial_v}) &:= \slashed{\delta}\mathbf{\bar{H}}_{\text{\tiny{NE}}}=\frac{1}{16\pi G}  \int_0^{2 \pi} \d \phi\, \left(-\Gamma\delta\bar{\Omega}_{{\text{\tiny{NE}}}}+\mathcal{U}\delta  \bar{\mathcal{J}}_{{\text{\tiny{NE}}}}+\mathcal{D}_{v}\bar{\Omega}_{{\text{\tiny{NE}}}}\delta\mathcal{P}_{{\text{\tiny{NE}}}} \right) \, .
\end{split}
\end{equation}
The zero mode charges for integrable non-expanding charges \eqref{integrable-Non-Integrable-charge-Y-NE} are 
\begin{equation}\label{Charges-zero-integrable-NE}
\begin{split}
        Q^{\text{I}}_{{\text{\tiny{NE}}}}({-r\partial_r}) &:= \frac{\mathbf{\bar{S}}_{{\text{\tiny{NE}}}}}{4\pi}=\frac{1}{16\pi G}  \int_0^{2 \pi} \d \phi\, \bar{\Omega}_{{\text{\tiny{NE}}}}\, ,     \\ 
        Q^{\text{I}}_{{\text{\tiny{NE}}}}({\partial_{\phi}}) &:= \mathbf{\bar{J}}_{\text{\tiny{NE}}}=\frac{1}{16\pi G}  \int_0^{2 \pi} \d \phi\, \bar{\mathcal{J}}_{{\text{\tiny{NE}}}}\, ,\\
       Q^{\text{I}}_{{\text{\tiny{NE}}}}({\partial_v}) &:= \mathbf{\bar{E}}_{\text{\tiny{NE}}}=\frac{1}{16\pi G}  \int_0^{2 \pi} \d \phi\, \left(\mathcal{U}{ \bar{\mathcal{J}}_{{\text{\tiny{NE}}}}}-\Gamma \bar{\Omega}_{{\text{\tiny{NE}}}}\right) \,.
\end{split}
\end{equation}
As it is clear from the above expressions all zero mode charges are different from the expanding cases.
\subsection{Null Boundary Thermodynamical Phase Space, Non-Expanding Case}\label{sec:Thermodynamical phase space-NE}
Similar to the expanding cases, here we present our general boundary thermodynamic picture for the non-expanding null surfaces.
\begin{enumerate}
\item[I.] Null boundary solution space for the non-expanding case similar to the VCT case consists of the following two parts: 
\begin{enumerate}
    \item[I.1)] thermodynamic sector: parametrized by $(\Gamma, {\cal U})$ and conjugate charges $(\bar{\Omega}_{\text{\tiny{NE}}}, \bar{\mathcal{J}}_{\text{\tiny{NE}}})$. They are subject to equations of motion \eqref{EOM-NE-1}.
    \item[I.2)] ${\mathcal{P}_{\text{\tiny{NE}}}}$, which only appears in  the flux and not in the integrable charge \eqref{integrable-Non-Integrable-charge-Y-NE}. Similar to the expanding cases, its thermodynamic conjugate is equal to the time derivative of the entropy aspect.
\end{enumerate}
    \item[II.] $\mathcal{P}_{\text{\tiny{NE}}}$ due to the entropy production takes our boundary system OTE.
    \item[III.] The time derivative of entropy aspect $\mathcal{D}_{v}\bar{\Omega}_{\text{\tiny{NE}}}$ measures the OTE from the boundary viewpoint.
\end{enumerate}
\subsection{Local First Law at Null Boundary}\label{subsec:first-law-NE}
Now, we are going to present a local first law for describing the thermodynamics of non-expanding null boundaries. Similar to the previous cases, we read the local first law from \eqref{charge-variation-NE-1} as follows
\begin{equation}\label{local-first-law}
\slashed{\delta} \bc{\bar{H}}_{\text{\tiny{NE}}}=T_{_{\mathcal{N}}} \delta\bc{\bar{S}}_{\text{\tiny{NE}}} +{\cal U}\delta\bc{\bar{J}}_{\text{\tiny{NE}}} +\frac{1}{\mu}\bc{\mathcal{T}}_{l\phi}\delta \bc{P}_{\text{\tiny{NE}}}
\end{equation}
where
\begin{equation}
    \bc{\bar{\mathcal{S}}}_{\text{\tiny{NE}}}=\frac{\bar{\Omega}_{\text{\tiny{NE}}}}{4 G}\, ,\hspace{1 cm}\bc{\bar{\mathcal{J}}_{\text{\tiny{NE}}}}=\frac{\bar{\mathcal{J}}_{\text{\tiny{NE}}}}{16\pi G}\, ,  \hspace{1 cm} \bc{P}_{\text{\tiny{NE}}}=\frac{\mathcal{P}_{\text{\tiny{NE}}}}{4\pi} \,.
\end{equation}
By using the equations of motion \eqref{EOM-NE-1}, we find 
\begin{equation}\label{local-first-law}
\boxed{\slashed{\delta} \bc{\bar{H}}_{\text{\tiny{NE}}}=T_{_{\mathcal{N}}} \delta\bc{\bar{S}}_{\text{\tiny{NE}}} +{\cal U}\delta\bc{\bar{J}}_{\text{\tiny{NE}}} +\mathcal{D}_{v}\bc{\bar{S}}_{\text{\tiny{NE}}}\delta \bc{P}_{\text{\tiny{NE}}}\, .}
\end{equation}
So, the local first law for the non-expanding case takes the same form as what we obtained in the VCT case \eqref{local-first-law-VCT} but now with different expressions for thermodynamic quantities.
\subsection{Local Gibbs-Duhem Equation at Null Boundary}\label{subsec:GD-relation-NE}
For the local Gibbs-Duhem equation, one gets
\begin{equation}\label{LEGD-eq-NE}
\boxed{\bc{\bar{E}}_{\text{\tiny{NE}}}=T_{_{\mathcal{N}}} \bc{\bar{S}}_{\text{\tiny{NE}}} +{\cal U} \bc{\bar{J}}_{\text{\tiny{NE}}}\, .}
\end{equation}
Again, the form of this local equation is similar to \eqref{LEGD-eq-VCT} but the explicit form of thermodynamic variables is different.
\subsection{Local Zeroth Law} \label{sec:Non-expanding-no-news}
For the non-expanding case, the on-shell variation of the action leads to
\begin{equation}\label{Onshell-deltaS-NE}
\begin{split}
    \delta I|_{\text{\tiny on-shell}}=\frac{1}{16\pi G} \int_{\mathcal{N}} \d v \d \phi \left(-\bar{\mathcal{J}}_{\text{\tiny{NE}}}\delta\mathcal{U}+\bar{\Omega}_{\text{\tiny{NE}}}\delta \Gamma+\frac{1}{\mu}\mathcal{T}_{l\phi}\delta\mathcal{P}_{\text{\tiny{NE}}}\right)\, .
    \end{split}
\end{equation}
On an equal footing with the previous expanding cases, to get a local thermal equilibrium, we require $ \delta I|_{\text{\tiny on-shell}}=\frac{1}{16\pi G}\int_{\mathcal{N}} \, \delta {\bar{\mathcal{G}}}_{\text{\tiny{NE}}}$ and hence we reach
\begin{equation}\label{Hamiltonian-NE}
    \delta\bc{\bar{\mathcal{H}}}_{\text{\tiny{NE}}}=\mathcal{U}\delta\bc{\bar{\mathcal{J}}}_{\text{\tiny{NE}}}+T_{_\mathcal{N}}\delta\bc{\bar{\mathcal{S}}}_{\text{\tiny{NE}}}+\frac{1}{\mu}\bc{\mathcal{T}}_{l\phi}\delta\bc{\mathcal{P}}_{\text{\tiny{NE}}}\, , \hspace{1 cm} \bc{\bar{\mathcal{H}}}_{\text{\tiny{NE}}}=\bc{\bar{\mathcal{G}}}_{\text{\tiny{NE}}}+T_{_\mathcal{N}}\bc{\bar{\mathcal{S}}}_{\text{\tiny{NE}}}+\mathcal{U}\bc{\bar{\mathcal{J}}}_{\text{\tiny{NE}}}\, .
\end{equation}
To have non-trivial solutions for this equation, we need to impose the following integrability conditions
\begin{equation}
    T_{_\mathcal{N}}\approx -\mathcal{D}_{v}\bc{\mathcal{P}}_{\text{\tiny{NE}}}=\frac{\delta\bc{\bar{\mathcal{H}}}_{\text{\tiny{NE}}}}{\delta\bc{\bar{\mathcal{S}}}_{\text{\tiny{NE}}}}, \hspace{1 cm} \frac{\bc{\mathcal{T}}_{l\phi}}{\mu}\approx \mathcal{D}_{v}\bc{\bar{\mathcal{S}}}_{\text{\tiny{NE}}}=\frac{\delta\bc{\bar{\mathcal{H}}}_{\text{\tiny{NE}}}}{\delta\bc{\mathcal{P}}_{\text{\tiny{NE}}}}, \hspace{1 cm} \mathcal{U}=\frac{\delta\bc{\bar{\mathcal{H}}}_{\text{\tiny{NE}}}}{\delta\bc{\bar{\mathcal{J}}}_{\text{\tiny{NE}}}}\, .
\end{equation}
Again these equations enforce an algebra among the surface charges
\begin{equation}\label{Charge-brackets-NE}
\begin{split}
  &\{\bc{\bar{S}}_{\text{\tiny{NE}}}(v,\phi), \bc{P}_{\text{\tiny{NE}}}(v,\phi')\}=\delta(\phi-\phi'),\quad  \{\bc{\bar{S}}_{\text{\tiny{NE}}}(v,\phi), \bc{\bar{S}}_{\text{\tiny{NE}}}(v,\phi')\}= \{\bc{P}_{\text{\tiny{NE}}}(v,\phi), \bc{P}_{\text{\tiny{NE}}}(v,\phi')\}=0,\\
&\{\bc{\bar{S}}_{\text{\tiny{NE}}}(v,\phi), \bc{\bar{J}}_{\text{\tiny{NE}}}(v,\phi')\}= \bc{\bar{S}}_{\text{\tiny{NE}}}(v,\phi'){\partial_{\phi}}\delta(\phi-\phi'), \\
&\{\bc{P}_{\text{\tiny{NE}}}(v,\phi), \bc{\bar{J}}_{\text{\tiny{NE}}}(v,\phi')\}= \left(\bc{P}_{\text{\tiny{NE}}}(v,\phi'){\partial_{\phi}}+ \bc{P}_{\text{\tiny{NE}}}(v,\phi){\partial_{\phi'}}+\frac{1}{\pi}\partial_{\phi'}\right)\delta(\phi-\phi'), \\
 &\{\bc{\bar{J}}_{\text{\tiny{NE}}}(v,\phi), \bc{\bar{J}}_{\text{\tiny{NE}}}(v,\phi')\}=\left(\bc{\bar{J}}_{\text{\tiny{NE}}}(v,\phi')\partial_{\phi}-\bc{\bar{J}}_{\text{\tiny{NE}}}(v,\phi)\partial_{\phi'}+\frac{1}{2\pi G\, \mu}\partial_{\phi'}^{3}\right)\delta(\phi-\phi')\, .
\end{split}
\end{equation}
This is the same as the charge algebra in the VCT case \eqref{Charge-brackets-VCT}, but with different coefficients for the central terms. 
\paragraph{Integrable slicing.}
Let us look at the following field dependent combinations of the symmetry generators
\begin{equation}
    \hat{W}=W-\Gamma T\, \hspace{1 cm} \hat{Y}=Y+\mathcal{U}T\, , \hspace{1 cm}  \hat{T}=\mathcal{D}_{v}\bar{\Omega}_{\text{\tiny{NE}}} T\, .
\end{equation}
Then, the charge variation
\begin{equation}\label{charge-variation-Y-1-NE}
    \begin{split}
        {\delta} Q_{\text{\tiny{NE}}}(\xi) \approx \frac{1}{16 \pi G} \int_0^{2 \pi} \d \phi \, (& \hat{W}\delta\bar{\Omega}_{\text{\tiny{NE}}}+\hat{Y}\delta\bar{\mathcal{J}}_{\text{\tiny{NE}}}+\hat{T}\delta\mathcal{P}_{\text{\tiny{NE}}})
    \end{split}
\end{equation}
will take an integrable form if we assume $\delta\hat{T}=\delta\hat{Y}=\delta\hat{W}=0$,
\begin{equation}\label{charge-variation-Y-1-NE}
    \begin{split}
        Q_{\text{\tiny{NE}}}(\xi) \approx \frac{1}{16 \pi G} \int_0^{2 \pi} \d \phi \, (& \hat{W}\bar{\Omega}_{\text{\tiny{NE}}}+\hat{Y}\bar{\mathcal{J}}_{\text{\tiny{NE}}}+\hat{T}\mathcal{P}_{\text{\tiny{NE}}})\, .
    \end{split}
\end{equation}
In this slicing, we get the same symmetry algebra as \eqref{Wb12-Tb12-Yb12} and the charge algebra also yields the following central term
\begin{equation}
    K^{\text{\tiny{NE}}}_{\xi_1,\xi_2}=\frac{1}{16\pi G}\int_{0}^{2\pi} \d {} \phi \left[\hat{T}_{2}\hat{W}_{1}+4\hat{Y}_{2}\partial_{\phi}\hat{T}_{1}+\frac{4}{\mu}\hat{Y}_{2}\partial_{\phi}^{3}\hat{Y}_{1}-(1\leftrightarrow 2)\right]\, .
\end{equation}
It is similar to \eqref{central-term-VCT}, but with different coefficients. This algebra with these coefficients matches with \eqref{Charge-brackets-NE}. 
\section{Outlook}\label{sec:discussion}
The main result in \cite{Adami:2021kvx} is that diffeomorphism invariance at the level of the action yields a local thermodynamic description for the boundary d.o.f. In this paper, we addressed the question of how essential is the diffeomorphism invariance to obtaining boundary thermodynamics. In this regard, we focused on theories with generally covariant equations of motion which are derived from actions which are generally invariant only up to some diffeomorphism non-invariant boundary terms. In particular we considered three dimensional topologically massive gravity theory. A careful analysis yields the fact that  diffeomorphism invariance at the level of  equations of motion is sufficient to guarantee a local thermodynamic description.

Specifically, we repeated an analysis similar to \cite{Adami:2021kvx} for the topologically massive gravity. We showed the boundary d.o.f which are labeled by the surface charges associated with large diffeomorphisms describe local boundary thermodynamics. We presented a local version of the first law, Gibbs-Duhem equation, and zeroth law which appear as a result of diffeomorphism invariance of equations of motion and account for the dynamics of part of spacetime behind the boundary. Our analyses extend the standard black hole thermodynamics in three different ways: 1) Our thermodynamic laws for the boundary system are local equations at the codimension 1 boundary. 2) Due to the passage of the bulk massive gravitons through the boundary and also the expansion of the boundary, our boundary system is out of thermal equilibrium. 3) Our analysis is true for any generic null boundary which need not be horizon of a black hole.

The VCT sector of solution phase space is parameterized by three surface charges. These surface charges describe local thermodynamics with local first law \eqref{local-first-law-VCT}, local Gibbs-Duhem \eqref{LEGD-eq-VCT}, and local zeroth law \eqref{zeroth-law-VCT-1}. As we discussed the latter  induces an algebra among the surface charges \eqref{Charge-brackets-VCT}. The form of these laws is exactly the same as what was obtained in \cite{Adami:2021kvx} for $D=3$ while the explicit form of thermodynamic variables receives contributions from the gravitational Chern-Simon term of TMG theory. We repeated these computations also for the full solution phase space NVCT, and obtained a set of local thermodynamic equations \eqref{local-first-law-NVCT-1}, \eqref{LEGD-eq-NVCT}, and \eqref{local-zeroth-law-NVCT-1}. 

In the NVCT case, in addition to the boundary d.o.f, we also have a massive bulk mode which through the interactions with the boundary d.o.f takes our boundary system out of thermal equilibrium. In this case because of the presence of news a comparison with the Einstein gravity in higher dimensions would be helpful. The form of the symplectic form \eqref{presymplectic} is comparable with what we have in the Einstein gravity in higher dimensions \cite{Adami:2021kvx}. The form of the local first law \eqref{local-first-law-NVCT-1} and local Gibbs-Duhem equation \eqref{LEGD-eq-NVCT}, except for two additional terms, are precisely in line with \cite{Adami:2021kvx}. These further terms carry the information about the underlying theory and hence we interpret them as the properties of TMG. Both of these terms are proportional to the news mode (graviton through the boundary) and do not contribute to the thermodynamics of stationary spacetimes. We postpone more detailed examination of them to  future work. Here we would like to discuss future projects and new directions.

\paragraph{Local second law.} In order to complete our null surface thermodynamics, we need to present a local version of the second law. Since null hypersurfaces are one-way membranes, we expect the exchanged energy and entropy from the boundary due to the passage of the flux of gravitons to have a definite sign. This property of null boundaries motivates us to present a local second law for our thermodynamic picture. In this regard, a careful analysis of the focusing theorem and the idea of light-sheets \cite{Bousso:1999xy,Bousso:2002ju} would be helpful. 

\paragraph{Relation to the membrane paradigm.} The common feature of what we presented here and the membrane paradigm \cite{PhysRevD.33.915,1986bhmp.book.....T} is the equivalence principle (diffeomorphism invariance of the theory). The general picture in the membrane paradigm is as follows: any observer which only has access to only the outside of the horizon should give a complete local account of physics without ever knowing what is inside the boundary. In other words, for this class of observers, one may excise the geometry at the horizon provided that we replace the excised region with a “membrane”  at the boundary of the excised region, the horizon. This picture is closely related to our thermodynamic picture. In the membrane paradigm, we project equations of motion on the horizon and try to interpret them in the hydrodynamics language. We believe our local thermodynamic equations are manifestations of this kind of boundary hydrodynamics. The main difference is that the membrane paradigm is based on the equations of motion but our local thermodynamics is based on the surface charge analysis. The balance equation is the main link between these two pictures.

\paragraph{Partially diffeomorphism invariant gravitational theories.} In this work, we considered TMG as a semi-covariant theory that does not have a diffeomorphism invariant action but the covariance appears at the level of the equations of motion. We have seen local thermodynamics describes the dynamics of the boundary d.o.f. One can go further and consider the actions with fewer diffeomorphisms. For example, in the Hořava–Lifshitz gravity \cite{Horava:2009uw} the time and space are not treated on an equal footing and hence we have only spatial diffeomorphism. An interesting question in this regard is whether one can associate a local thermodynamic description for the boundary d.o.f in these kinds of theories with a part of diffeomorphisms.

\paragraph{Higher curvature theories.} As we mentioned the local description of the boundary d.o.f is a result of the diffeomorphism invariance of the bulk theory. It is worth doing the same analysis for higher curvature theories and driving similar local equations for the first law, Gibbs-Duhem equation, and the zeroth law. We conjecture for expanding null surfaces, the surface charge associated with supertranslation, in the integrable slicing, does not admit any corrections from the higher curvature terms.  Furthermore, the chemical potential associated with this charge is always equal to the covariant derivative of the null surface entropy. Examination of this conjecture would be interesting and it may shed light on dynamic processes such as the black hole formation and evaporation.
\paragraph{Acknowledgement.}
The author is greatly indebted to Shahin Sheikh-Jabbari and Mohammad Hassan Vahidinia for illuminating discussions. He also would like to especially thank Hamed Adami, Hossein Yavartanoo, Daniel Grumiller, Celine Zwikel and Pujian Mao for long term collaborations and many fruitful discussions which were crucial in developing the ideas and analysis discussed here. This work has been partially supported by IPM funds and also the Grant of National Elites Foundation of Iran.

\appendix

\bibliographystyle{fullsort.bst}
\bibliography{reference}

\end{document}